\documentclass[prb,twocolumn,showpacs]{revtex4}
\usepackage{graphicx,amsmath,epsf}
\DeclareMathOperator{\trace}{tr}

\DeclareMathOperator{\cov}{cov}

\newcommand{\vv}{\mathbf{v}}
\newcommand{\vR}{\mathbf{R}}

\begin{document}
\title{Classical limit of transport in quantum kicked maps}

\author{Saar Rahav}
\affiliation{Laboratory of Atomic and Solid State Physics, Cornell University, Ithaca 14853, USA.}
\author{Piet W. Brouwer}
\affiliation{Laboratory of Atomic and Solid State Physics, Cornell University, Ithaca 14853, USA.}

\begin{abstract}
We investigate the behavior of weak localization, conductance fluctuations,
and shot noise of a chaotic scatterer in the semiclassical limit.
Time resolved numerical results, obtained by truncating the 
time-evolution of a kicked quantum map after a certain number of
iterations, are compared to semiclassical theory. Considering how
the appearance of quantum effects is delayed as a function of the 
Ehrenfest time gives a new method to compare theory and numerical
simulations. We find that both weak localization and
shot noise agree with semiclassical theory, which
predicts exponential suppression with increasing Ehrenfest time. 
However, conductance fluctuations exhibit different behavior, with 
only a slight dependence on the Ehrenfest time. 
\end{abstract}

\maketitle

\section{Introduction}

According to Ehrenfest's theorem, the expectation values of the 
position and momentum of an electron obey classical 
equations of motion. As long as the wavefunction of the electron 
is a wavepacket with minimal uncertainties in momentum and position,
the expectation values are a good description of the quantum state.
However, wavepackets disperse, and the Ehrenfest theorem
looses its relevance after a short time. In a cavity with
point scatterers, which split electron wavepackets into partial
waves after one scattering event, this time is simply the elastic 
mean free time. In a ballistic cavity with chaotic classical dynamics, 
this time is the so-called
``Ehrenfest time'' $\tau_{\rm E}$, which depends on
the Lyapunov exponent $\lambda$ of the classical motion in the
cavity.\cite{kn:zaslavsky1981,kn:aleiner1996}
For times longer
than $\tau_{\rm E}$, a classical description no longer holds and
the wave nature of the electrons becomes visible.

The wave nature of electrons is the cause of some striking effects 
that are
absent in classical systems. For transport through cavities 
coupled to source and drain reservoirs via point contacts, 
these effects are weak localization, universal
conductance fluctuations, and shot noise.\cite{kn:beenakker1991b} 
In the limit that transport through cavities is ergodic (dwell time
in the cavity is much longer than the time of flight through the
cavity), the signatures of quantum transport are `universal', 
independent of the cavity size and shape, and of the 
fact whether electron motion inside the cavity is ballistic and
chaotic or diffusive, with repeated scattering off impurities with
size smaller than the electron wavelength. Random matrix theory 
provides a unified theoretical description of weak localization, 
universal conductance fluctuations, and shot noise in ballistic
or diffusive cavities.\cite{kn:beenakker1997}

If the electron motion is diffusive, the dynamics is fully quantum 
mechanical already at times much shorter than the time $\tau_{\rm
  erg}$ required for ergodic exploration of the cavity's phase
space. For ballistic cavities this is
true in most practical applications as well
--- the Ehrenfest time $\tau_{\rm
E}$ usually does not exceed the time of flight through the cavity ---
but there is no fundamental reason why $\tau_{\rm E}$ always has to be
small. The case of large Ehrenfest times is of theoretical interest,
as it is one of very few regimes in parameter space in which one can
observe differences between signatures of quantum transport in
ballistic chaotic and diffusive cavities.

The most prominent effects of a large Ehrenfest time are found if
$\tau_{\rm E}$ is larger than the dwell time $\tau_{\rm D}$ in 
the cavity. If $\tau_{\rm E} \gg \tau_{\rm D}$, quantum transport
is deterministic, and shot noise is
suppressed.\cite{kn:beenakker1991c,kn:agam2000} The suppression
of shot noise has been observed experimentally by varying the dwell
time $\tau_{\rm D}$ of a chaotic cavity,\cite{kn:oberholzer2002} 
and numerically, using a chaotic map as a model for a chaotic
cavity.\cite{kn:tworzydlo2003,kn:jacquod2004,kn:jacquod2005}
The effect of a large Ehrenfest time on weak localization was first 
addressed by Aleiner and Larkin.\cite{kn:aleiner1996} Their theory
predicts a suppression of weak localization $\propto \exp(-
\tau_{\rm E}/\tau_{\rm D})$, if classical correlations are taken 
into account properly.\cite{kn:rahav2005} The same suppression was
found in an independent calculation by Adagideli.\cite{kn:adagideli2003}
Experimental observation
of the suppression of weak localization at large Ehrenfest times
has been reported for transport through antidot 
arrays.\cite{kn:yevtushenko2000} No semiclassical theory for the
Ehrenfest-time dependence of universal conductance fluctuations
exists. However, semiclassical theories for weak localization and
universal conductance fluctuations for the limit $\tau_{\rm D}
\gg \tau_{\rm E}$ are essentially 
equal,\cite{kn:argaman1995,kn:argaman1996,kn:takane1997,kn:takane1998} 
as are diagrammatic perturbation theories for the same phenomena in 
diffusive cavities, supporting the expectation that
the Ehrenfest-time dependencies of weak localization and universal
conductance fluctuations will be equal as well.\cite{kn:tworzydlo2004c}

Direct numerical simulation of the effect of a large Ehrenfest 
time on quantum transport through two-dimensional
chaotic cavities has been problematic because of the 
prohibitively high computational cost of the simulations.
The reason is that $\tau_{\rm E}$ depends only logarithmically
on the product of the electron wavenumber $k$ and the cavity size
$L$,
\begin{equation}
  \tau_{\rm E} = \lambda^{-1} \ln k L.
\end{equation}
For two-dimensional cavities, system sizes at
which $\tau_{\rm E} \gtrsim \tau_{\rm D}$ cannot be simulated with
present-day algorithms and processor speeds.
In order to circumvent this problem, Jacquod,
Schomerus, and Beenakker proposed to replace the cavity by a
quantum map.\cite{kn:jacquod2003} The map is 'opened', so that
simulation of transport properties is possible.
Although a map has a one-dimensional
phase space, a chaotic map shares many characteristics of the
chaotic motion in two-dimensional chaotic
cavities.\cite{kn:stoeckmann1999} The reduced dimensionality of
the map's phase space made numerical simulations with larger
Ehrenfest times possible. For an open version of the quantum 
kicked rotator map,
numerical simulations were reported for shot
noise,\cite{kn:tworzydlo2003,kn:jacquod2004,kn:jacquod2005}
weak localization,\cite{kn:tworzydlo2004b,kn:rahav2005} and universal
conductance 
fluctuations.\cite{kn:jacquod2004,kn:tworzydlo2004,kn:tworzydlo2004c}
Simulation results for shot noise were in good agreement
with the predictions of the semiclassical theory.\cite{kn:agam2000}
However, for conductance fluctuations, no dependence 
on $\tau_{\rm E}$ was found, despite the fact that Ehrenfest times
larger than the dwell time were considered.~\cite{footnote}
Whereas early numerical simulations of weak localization showed no 
Ehrenfest-time dependence,\cite{kn:tworzydlo2004b} we showed that 
there is a
systematic decrease of the weak localization correction to the 
conductance upon increasing $\tau_{\rm E}$, consistent with the
semiclassical theory.\cite{kn:rahav2005}

\begin{figure}
\epsfxsize=0.8\hsize
\epsffile{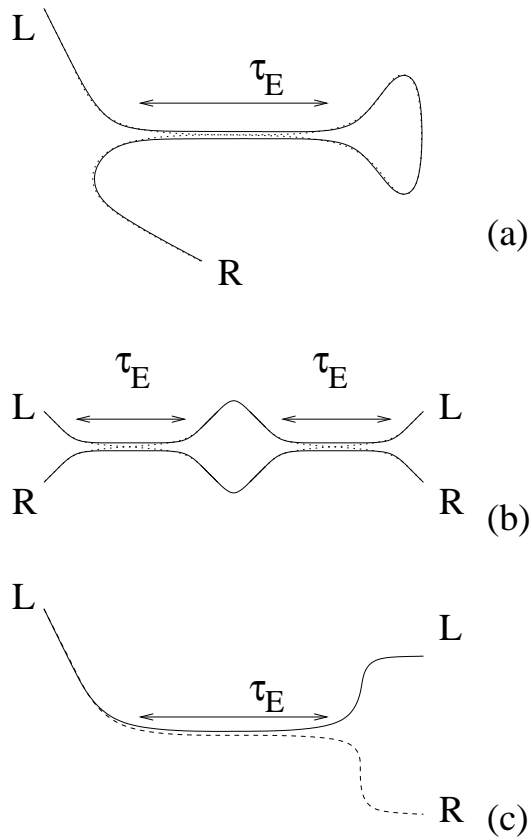}
\caption{\label{fig:1} Schematic drawing of relevant trajectories
for weak localization (a), conductance fluctuations (b), and
shot noise (c). The letters ``L'' and ``R'' refer to the left and
right contacts to the cavity, respectively. The trajectories shown
in panel (a) are for the weak localization to the transmission $T$;
the trajectories shown in panel (b) are for the covariance 
$\mbox{cov}(R_{\rm L},R_{\rm R})$ of reflections from the left and 
right contacts. Both weak localization and conductance fluctuations
require a minimal dwell time of $2 \tau_{\rm E}$. Shot noise requires
a minimal dwell time $\tau_{\rm E}$ only.}
\end{figure}

The main technical innovation that allowed us to detect a
systematic decrease of the weak localization correction upon
increasing the Ehrenfest time is that we looked at time-resolved
numerical simulations: The map's time evolution is truncated after
a time $t_0$, and weak localization, conductance fluctuations, and
shot noise are monitored as a function of $t_0$.\cite{kn:rahav2005}
This procedure has two advantages. First, it allows the ensemble
average over the quasienergy $\varepsilon$ to be done analytically. 
(See Sec.\ \ref{numerics} for technical details.)
This made it possible to consider significantly larger ensembles 
than  considered previously.
Second, monitoring quantum corrections as a function of the
`truncation time' $t_0$ allows us to determine the minimal time
after which quantum corrections can occur. 
In the semiclassical theory, quantum interference 
requires a minimal wavepacket to be split {\em and} reunited,
which takes a minimal time $2 \tau_{\rm E}$. A schematic diagram
drawing relevant semiclassical trajectories for weak localization
and conductance fluctuations is shown in Fig.\ \ref{fig:1}. (The
diagram for conductance fluctuations is taken from Ref.\
\onlinecite{kn:takane1998} and modified to contain the effect
of a finite Ehrenfest time.) Not
being a quantum interference effect, shot noise only requires 
wavepackets to be split, which happens after a time $\tau_{\rm E}$.
Comparison of the time when quantum effects appear (the `onset time')
and the rate of suppression of quantum effects as the Ehrenfest 
time $\tau_{\rm E}$
is increased, thus provides a quantitative test of the semiclassical
theory. With such a quantitative test of the semiclassical theory, 
accurate simulations performed at smaller Ehrenfest times can still be
meaningful.

In this paper, we present detailed results of such time-resolved 
numerical simulation for weak localization, conductance fluctuations,
and shot noise. For all three quantum effects, we analyze their
Ehrenfest-time dependence (without truncation of the time evolution)
and the onset times (obtained from simulations with truncation of
the time evolution). The simulation results are presented in Sec.\
\ref{kicked}, together with the predictions of random matrix theory
for time-resolved transport through open quantum maps. Whereas we
confirm our earlier conclusion that the numerical simulations for
weak localization show a suppression $\propto \exp(-\tau_{\rm
  E}/\tau_{\rm D})$, consistent with the semiclassical theory, we
find that simulations for conductance fluctuations show a very
small increase if $\tau_{\rm E}$ is increased, the effect being
small enough to be consistent with the 
simulation data reported in the
literature.\cite{kn:jacquod2004,kn:tworzydlo2004,kn:tworzydlo2004c}
The onset times for conductance fluctuations are more than a factor two
smaller than the onset times for weak localization, which is 
incompatible with the notion that conductance fluctuations arise
as wavepackets are split and reunited. 

A second goal of this paper is to show how classical correlations
are taken into account in the semiclassical theory of 
Ref.\ \onlinecite{kn:aleiner1996}. The importance of classical
correlations --- no quantum diffraction takes place to or from 
classical trajectories with classical dwell time shorter than
the Ehrenfest time --- was pointed out in the analysis of
simulation data for the quantum kicked rotator, most notably the
simulations for shot noise.\cite{kn:tworzydlo2003} The original
version of the semiclassical theory,\cite{kn:aleiner1996} which 
did not include these correlations, predicted a suppression of
weak localization $\propto \exp(-2 \tau_{\rm E}/\tau_{\rm D})$.
It is only after accounting for the classical correlations 
that the proper exponential decay $\propto \exp(-\tau_{\rm
  E}/\tau_{\rm D})$ is recovered. (Classical correlations are
taken into account correctly in the semiclassical theory of 
shot-noise suppression at large Ehrenfest times.\cite{kn:agam2000})
In addition it is demonstrated that this semiclassical 
theory is unitary, {\em i.e.,} that no probability is lost.
Our discussion of the semiclassical theory for weak localization
can be found in Sec.\ \ref{ALcalculation}. 
We conclude in Sec.\ \ref{discussion}.

\section{Time resolved transport in open quantum maps}
\label{kicked}
\label{numerics}

Despite their different dimensionality,
chaotic quantum maps are believed to have the same phemonology as
(closed) chaotic 
cavities,\cite{kn:fishman1982,kn:altland1996,kn:borgonovi1996,kn:frahm1997}
provided one restricts attention to the ``universal limit'' of times
longer than the ``ergodic time''. (The ``ergodic time'' is the time
required to explore the phase space.)
Open kicked quantum maps have been used as a model for transport 
through chaotic cavities,\cite{kn:jacquod2003} because maps
allow to simulate systems with many transport channels
with relatively small computational effort. For short times,
qualitative differences between maps and cavities exist, however,
because the time evolution in maps is discrete, whereas the time
evolution in cavities is not.

Quantum maps operate on a finite state
vector $\psi$ of dimension $M$. In the context of the quantum
kicked rotator these states are discrete quasimomentum states or, 
alternatively, positions
on a lattice with periodic boundary conditions. In the language of
a cavity, the elements of $\psi$ can be thought of as points on 
the cavity boundary. The time evolution of such maps is discrete,
and given by the Floquet operator
\begin{equation}
\psi (t+1) = {\cal F}\psi (t).
\end{equation} 
For the kicked rotator, this type of dynamics is relatively easy 
to simulate numerically, which is why it is used to numerically
model quantum interference corrections at large Ehrenfest times.

In order to study transport, the system has to be opened. Hereto
two consecutive sets of $N_{\rm L}$ and $N_{\rm R}$ elements 
$\psi$ are chosen that
corresponds to two `leads'.\cite{kn:jacquod2003} 
The initial condition for a transport
simulation corresponds to $\psi$ localized at one of the `lead'
points. Escape from the cavity is modeled by recording the amplitude
of $\psi$ at the lead points at time $t$ and setting $\psi$ to 
zero afterwards. Formally, this corresponds to the 
construction\cite{kn:brouwer1999a,kn:fyodorov2000,kn:ossipov2003,kn:tworzydlo2003} 
\begin{equation}
\label{seps}
S(\varepsilon) = P \left[ 1 - e^{i \varepsilon}
  {\cal F} Q \right]^{-1} e^{i \varepsilon} {\cal F} P^T,
\end{equation}
where $S$ is the scattering matrix corresponding to the map, 
$\varepsilon$ is quasienergy, $P$ is a $(N_{\rm R} +
N_{\rm L}) \times M$ 
matrix projecting on the sites corresponding to the left (L) 
and right (R) contacts, and $Q = 1 - P^{T} P$. 
The conductance coefficients $G_{\rm LL}$, $G_{\rm RL}$, $G_{\rm LR}$,
and $G_{\rm RR}$ are defined as
\begin{eqnarray}
\label{conductances}
  G_{\rm RL} & \equiv& \mbox{tr}\, S C_{\rm L}
  S^{\dagger} C_{\rm R}, \nonumber
  \\
  G_{\rm LR} & \equiv& \mbox{tr}\, S C_{\rm R}
  S^{\dagger} C_{\rm L} \nonumber
  \\
  G_{\rm LL} &\equiv& \mbox{tr}\, S C_{\rm L}
  S^{\dagger} C_{\rm L} - N_{\rm L}, \nonumber
  \\
  G_{\rm RR} &\equiv& \mbox{tr}\, S C_{\rm R}
  S^{\dagger} C_{\rm R} - N_{\rm R},
\end{eqnarray}
where $C_{\rm R}$ projects on the channels of the right lead,
whereas $C_{\rm L}$ projects on the channels of the left lead.
Unitarity implies $G_{\rm RL} = G_{\rm LR} = - G_{\rm LL} = 
- G_{\rm RR}$.
The dwell time corresponding to the map is $\tau_{\rm D} = M/(N_{\rm
  L} + N_{\rm R})$.

We note that, with this method of opening the map, the leads end 
abruptly. Such abrupt changes lead to diffraction effects, similar 
to diffraction from sharp corners at the lead opening
of a cavity. The semiclassical contributions of some diffracting
orbits of this type were calculated for closed and open cavities,
see, {\em e.g.,} Refs.\
\onlinecite{kn:sieber1997,kn:bogomolny2000,kn:stampfer2005}. They were
found to be an important contribution to shot noise at small channel
numbers $N$ for rectangular
cavities.\cite{kn:aigner2005,kn:marconcini2005}
However, to 
the best of our knowledge, the total
contribution of such orbits to shot noise, weak localization,
and conductance fluctuations in chaotic cavities at large $N$ are 
unknown. Therefore, we compare
our numerical results to the existing semiclassical theory,
which neglects the effects of corner diffraction.

In this section, we will study the transmission and reflection
coefficients as a function of a truncation time $t_{0}$. The
truncation procedure involves writing Eq.\ (\ref{seps}) as a
geometric series of a time-dependent scattering matrix $S(t)$,
\begin{equation}
\label{sexpandt}
S(\varepsilon) = \sum_{t=1}^\infty e^{i \varepsilon t} S(t),
\end{equation}
with 
\begin{equation}
\label{soft}
S (t) = P \left[ {\cal F} Q \right]^{t-1} {\cal F} P^T.
\end{equation}
Instead of taking the full geometric series (\ref{sexpandt}),
we truncate the series after time $t_0$,
\begin{equation}
S(t \le t_0;\varepsilon) \equiv \sum_{t=1}^{t_0} e^{i \varepsilon t}
S(t),
  \label{eq:Strunc}
\end{equation}
and study transport properties as a function of $t_0$. Although
such a truncation procedure does not represent a physical system,
it allows us to get theoretical information on the times involved
in quantum transport phenomena. In particular, we will verify at
what truncation time weak localization and conductance fluctuations 
will first appear.

The quantities of interest are the ensemble-averaged conductance
coefficients $G_{\rm LL}$, $G_{\rm RL}$, $G_{\rm LR}$, and 
$G_{\rm RR}$, with $S$ replaced by $S(t \le t_0)$,  and the 
fluctuations of these coefficients. With the truncation procedure,
$S$ is no longer unitary, so that 
one does not necessarily have that $G_{\rm RL} = G_{\rm LR} = - G_{\rm
  LL} = - G_{\rm RR}$. Hence, we need to consider all four conductance
coefficients separately.
If ${\cal F}$ is time-reversal symmetric, one still has $G_{\rm RL} 
= G_{\rm LR}$, and only three conductance coefficients need to be
considered. For $t_0 \gg \tau_{\rm D}$, unitarity is restored, and
one conductance coefficient is sufficient.

It is also of interest to study the $t_0$ dependence of shot noise.
This also can be done by replacing the scattering matrix by $S(t \le t_0)$, as
was done for weak localization and conductance fluctuations.
The resulting $t_0$-dependent Fano factor is given by~\cite{kn:tworzydlo2003}
\begin{equation}
\label{eq:fano}
F (t_0) = \frac{N_{\rm R}+N_{\rm L}}{N_{\rm R} N_{\rm L}} \trace \left[S C_{\rm R} S^\dagger C_{\rm L} \left(1 - S C_{\rm R} S^\dagger C_{\rm L}\right) \right].
\end{equation}

The
ensemble average is taken by first averaging over the quasienergy
$\varepsilon$, and 
then over various lead positions. The quasienergy average can be
performed explicitly, see Eq.\ (\ref{eq:Strunc}) above. For the
fluctuations, we consider the variances of the transport
coefficients, as well as covariances of different coefficients.
The variances are taken with respect to the quasienergy $\varepsilon$,
in order to ensure that the fluctuations are entirely of quantum
origin.\cite{kn:tworzydlo2004,kn:jacquod2004} 
The variance or covariance for fluctuations with respect 
to $\varepsilon$ is then averaged over different lead positions.
Again, having the explicit energy dependence (\ref{eq:Strunc}) at
our disposal, the calculation of the variance or covariance with 
respect to variations of $\varepsilon$ can be performed explicitly.

\subsection{Random matrix theory of time-resolved transport}

As a reference for our numerical simulations, in which we take
the Floquet operator ${\cal F}$ of the quantum kicked rotator, we 
consider averages and fluctuations of the transport coefficients
for the case that ${\cal F}$ is a random symmetric unitary matrix,
taken from the circular orthogonal ensemble of random matrix theory.
In the limit of 
large $M$, $N_{\rm L}$, and $N_{\rm R}$, which is relevant for
the semiclassical limit we consider throughout this paper,
such averages can be
calculated using the technique of Ref.\ \onlinecite{kn:brouwer1996a}.
We then find 
\begin{equation}
  \langle G_{\alpha \beta}(t \le t_0) \rangle
  = \langle G_{\alpha \beta}(t \le t_0) \rangle_{\rm cl} + 
  \delta G_{\alpha \beta}(t \le t_0)+O(M^{-2}),
\end{equation}
where the indices $\alpha$ and $\beta$ can be taken to be L and 
R. The first term $\langle G_{\alpha \beta}(t_0) \rangle_{\rm cl}$ is the
classical average,
\begin{equation}
  \langle G_{\alpha \beta}(t \le t_0)\rangle _{\rm cl} =
  \frac{N_{\alpha} N_{\beta}}{M}
  \frac{1 - x^{t_0}}{1 - x}
  - N_{\alpha} \delta_{\alpha \beta},
\end{equation}
with $x=1 - 1/\tau_{\rm D}$, 
and $\delta G_{\alpha \beta}(t_0)$ is the weak localization correction,
\begin{widetext}
\begin{eqnarray}
  \delta G_{\alpha \beta}(t \le t_0) &=&
  \frac{N_{\alpha} (1 - x^{t_0})}{M(1-x)} \delta_{\alpha \beta}
  - \frac{N_{\alpha} N_{\beta}}{M^2}
  \left( \frac{1 - x^{t_0}}{(1-x)^2}
  - \frac{t_0 x^{t_0}}{1 - x}
  + \frac{t_0 (t_0-1) x^{t_0-1}}{2} \right).
\end{eqnarray}
For the covariances we find similarly in the limit of $M$, $N_{\rm
  R}$, $N_{\rm L} \gg 1$,
\begin{multline}
\label{covariances}
  \langle \cov_{\varepsilon} \left[ G_{\alpha \beta} (t \le t_0) ,
    G_{\gamma \delta} (t \le t_0) \right] \rangle
  = 2 \frac{N_\alpha N_\beta N_\gamma N_\delta}{(N_{\rm L} + N_{\rm R}+1)^4} {\cal R}_1 + 2 \frac{N_\alpha N_\beta}{(N_{\rm R}+N_{\rm L}+1)^3} \left( N_\gamma \delta_{\beta \delta} + N_\gamma \delta_{\alpha \delta} + N_\delta \delta_{\alpha \gamma} + N_\delta \delta_{\beta \gamma} \right) {\cal R}_2  \\  +  2 \frac{N_\alpha N_\beta}{(N_{\rm R}+N_{\rm L}+1)^2} \left( \delta_{\alpha \gamma} \delta_{\beta \delta} + \delta_{\alpha \delta} \delta_{\beta \gamma} \right) {\cal R}_3+O(M^{-1}),
\end{multline}
where
\begin{align}
{\cal R}_1 & =  \frac{2x}{1+x} -2 x^{t_0-1} \left( (t_0+1) x^2 +6x -t_0+1 \right) - \frac{2 x^{2t_0-1}}{1+x} +  x^{2t_0-2} \nonumber \\ & \times  \left( - \frac{2}{3} (1-x)^3 t_0^3 + (1-x)^2 (1+x) t_0^2 + \frac{1}{3} t_0 (1-x)(11x^2 + 20x -1) + 2x (x^2 + 5x +2) \right), \nonumber \\
{\cal R}_2 & =  - \frac{x}{1+x} +2 x^{t_0} + \frac{x^{2t_0 -1}}{1+x} - (t_0+1) x^{2 t_0 -1} + (t_0 -1)x^{2 t_0} , \nonumber \\
{\cal R}_3 & =  x \left( \frac{1}{1+x} -x^{t_0-1} +\frac{x^{2 t_0 -1}}{1+x} \right).
\end{align}
\end{widetext}
In Eq.\ (\ref{covariances}) the symbol $\mbox{cov}_{\varepsilon}$
denotes a covariance taken with respect to variations of the
quasienergy only, whereas the brackets $\langle \ldots \rangle$ denote
the ensemble average over ${\cal F}$.

\subsection{The open quantum kicked rotator}

In our numerical simulations, we take the Floquet operator ${\cal F}$
of the quantum kicked rotator map. The map is described in detail by
Tworzydlo {\em et al.}, see Ref.\
\onlinecite{kn:tworzydlo2004b}. The matrix elements
of the Floquet operator are
\begin{equation}
\label{onekick}
{\cal F}_{mn} = \left( X U^{\dagger} \Pi U X \right)_{mn}
\end{equation}
with
\begin{align*}
  U_{mn} & =  M^{-1/2} e^{2 \pi i m n /M},  \\
  X_{mn} & =  \delta_{mn} e^{-i \frac{MK}{4 \pi} 
    \cos \left( 2 \pi m/M+\phi \right)},  \\
  \Pi_{mn} & =  \delta_{mn} e^{-i \pi m^2/M}. 
\end{align*}
Here $K$ is the so called stochasticity parameter that determines the classical
dynamics of the map. The region $K \gtrsim 7.5$ is associated with
classically chaotic dynamics. The parameter $\phi$ determines the
precise quantization of the map and has no effect on the classical
dynamics. In this model the size of the matrix $M$ is even. In the 
simulations we set $N_{\rm R} = N_{\rm L} = N$.

We used the Floquet operator (\ref{onekick}) to study conductance
fluctuations and shot noise. The Floquet operator (\ref{onekick})
can also be used to study weak localization using the following
argument: The average conductance $\langle G_{\alpha \beta} \rangle$ consists of
a classical contribution and a quantum correction,
\begin{equation}
  \langle G_{\alpha \beta} \rangle = G_{\alpha \beta,{\rm cl}} + \delta G_{\alpha \beta}.
\end{equation}
The classical contribution $G_{\alpha \beta,{\rm cl}}$ scales proportional to the 
channel number $N$, whereas the quantum correction has no
$N$-dependence (except for a possible weak dependence on $N$ through 
the Ehrenfest time). The quantum correction can be extracted by 
comparing average conductances at $N$ channels and $2N$ channels,
\begin{eqnarray}
  \delta G_{\alpha \beta} &\approx& 2 \delta G_{\alpha \beta}(N) - \delta G_{\alpha \beta} (2 N) \nonumber \\
  &=& 2 \langle G_{\alpha \beta} (N) \rangle - \langle G_{\alpha \beta} (2N) \rangle.
  \label{eq:scaling}
\end{eqnarray}
In order to avoid a spurious contribution to $\delta G_{\alpha \beta}$ from
classical conductance fluctuations, we made sure that the
ensemble averages for $\langle G_{\alpha \beta} (2N) \rangle$ and $\langle
G_{\alpha \beta} (N)\rangle$ were taken for {\em precisely} the same classical
dynamics (same values of $K$, same lead positions). This method
was used previously in Ref.\ \onlinecite{kn:rahav2005}.

Alternatively, one can study weak localization by considering
maps with and without time reversal symmetry. 
The kicked rotator map (\ref{onekick}) has time-reversal symmetry. 
A simple extension of
Eq.\ (\ref{onekick}) that breaks time-reversal symmetry is the so-called 
three-kick model,\cite{kn:tworzydlo2004b}
\begin{equation}
\label{3kick}
{\cal F}_{mn} = \left(X \Pi Y^* \Pi Y \Pi X \right)_{mn}
\end{equation}
where
\begin{align*}
Y_{mn} & =  \delta_{mn} e^{i \frac{\gamma M}{6 \pi} \cos \left(2 \pi m/M\right)}, \\ 
X_{mn} & =  \delta_{mn} e^{-i \frac{M}{12 \pi} V(2 \pi m/M)}, \\
\Pi_{mn} & =  M^{-1/2} e^{-i \pi/4} e^{i \frac{\pi}{M} (n-m)^2}. 
\end{align*}
In this model $M$ is even, but not a multiple of $3$. The kick potential is
given by
\begin{equation}
V (\theta) = K \cos \left( \pi q/2 \right) \cos \theta + \frac{1}{2} K \sin  \left( \pi q/2 \right) \sin 2\theta,
\end{equation}
where $q$ breaks the parity symmetry of the model.\cite{kn:bluemel1992b} The parameter
$\gamma$ plays the role of a magnetic field and breaks time-reversal 
symmetry.  By
comparing the transport for $\gamma \ne 0$ to that of $\gamma=0$ it is
possible to investigate the weak localization correction as a function
of time. This is the method used in Ref.\ \onlinecite{kn:tworzydlo2004b}.

The advantage of the second method is that it does not require the
cancellation of classical conductances of two different systems. The
disadvantage is that it involves the additional parameter $\gamma$, 
which itself also affects the map's classical dynamics. However,
weak localization is affected on the scale $\gamma \sim 1/N$,
so that the effect on the classical dynamics can be expected to
be small in the semiclassical limit $N \gg 1$.

For both models, the Ehrenfest time is given by
\begin{equation}
  \tau_{\rm E} = \lambda^{-1} \ln N, \label{eq:tEmap}
\end{equation}
up to an $N$-independent constant. The dwell time
reads
\begin{equation}
  \tau_{\rm D} = \frac{M}{2 N}.
\end{equation}

\subsection{Numerical results}

The numerical algorithm used
to iterate the maps (\ref{onekick}) and (\ref{3kick}) can be accelerated
by using a fast Fourier
transform.\cite{kn:tworzydlo2004,kn:tworzydlo2003,kn:tworzydlo2004b,kn:ketzmerick1999}
The results are computed as a function of time and not of quasienergy.
This allows to truncate the series (\ref{sexpandt})
at the maximal time studied,
and to calculate the $\varepsilon$-averages explicitly.

\begin{figure}
\epsfxsize=0.9\hsize
\epsffile{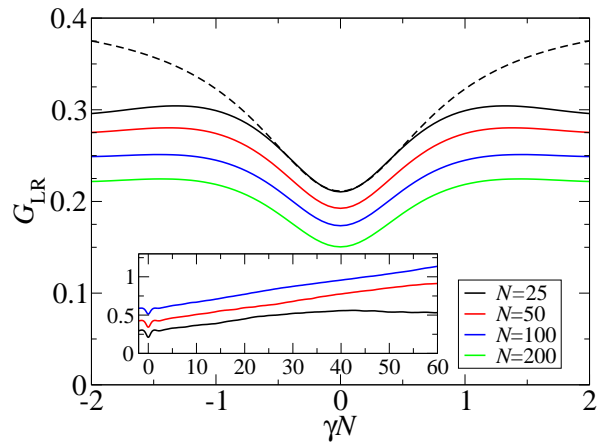}
\caption{\label{fig:wl1} Average conductance for the three-kick
model (\protect\ref{3kick}) as a function of the time-reversal
symmetry breaking parameter $\gamma N$. Curves shown for
$N=25,50,100$ and $200$ are offset vertically. The stochasticity
parameter $K \approx 10$, $q=0.2$, and the dwell time $\tau_{\rm D} = 5$. 
Inset: Same for a larger range of $\gamma N$ values.}
\end{figure}

{\em Weak localization}.
The value of the time-reversal symmetry breaking parameter 
$\gamma$ at which weak localization corrections to the conductance
are suppressed in the three-kick model (\ref{3kick}) scales 
inversely proportional to the channel number
$N$.\cite{kn:tworzydlo2004b} Hence, in order to compare weak 
localization corrections for different channel numbers, the
average conductance is calculated as a function of the product
$\gamma N$. Figure \ref{fig:wl1} shows the result of numerical
simulations of the ensemble-averaged conductance. The 
ensemble average is taken over 1000 samples, choosing $K$ 
randomly in the interval $10 < K < 11.5$ and varying the lead
positions. The figure also shows a Lorentzian fit to the simulation
data for small $\gamma N$. 
The dwell time is set to be $\tau_{\rm D} = 5$ and the
parity-symmetry breaking parameter $q = 0.2$.
Results for different $N$ are offset vertically. 
When determining the magnitude of the weak localization correction, we
restrict our attention to the range $\gamma N \lesssim 1$, for which
the $\gamma N$-dependence of $\langle G_{\rm LR} \rangle$ exhibit a pronounced
dip. The width of this dip seems to be (almost) independent on
system size, in agreement with the predictions from random matrix
theory.\cite{kn:tworzydlo2004b}
However, as can be seen from Fig. \ref{fig:wl1}, the shape of this 
peak is not well approximated by a Lorentzian at higher 
values of $\gamma$.
For $\gamma N \gtrsim 1$ the average conductance typically continues to
increase with $\gamma N$, but at a much slower rate.

\begin{figure}
\epsfxsize=0.9\hsize
\epsffile{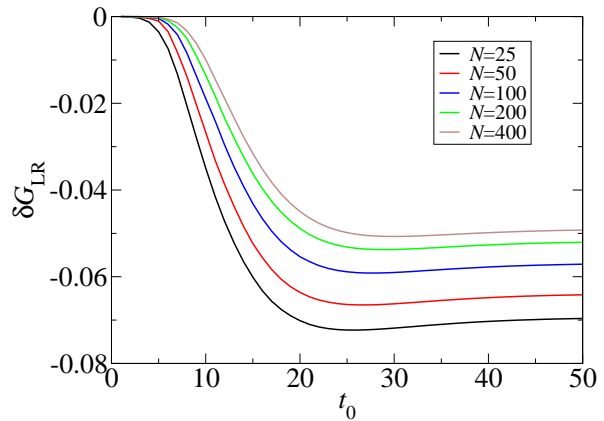}
\caption{\label{fig:wl2} Time-resolved 
difference between ensemble-averaged
conductance at $\gamma N = 0$ and $\gamma N = 0.7$ for the three-kick
model (\protect\ref{3kick}). The dwell time is $\tau_{\rm D} =5$, while
$K \approx 10$ and $q=0.2$}
\end{figure}

Figure \ref{fig:wl2} shows the difference $\delta G_{\rm LR} (\gamma) = \langle
G_{\rm LR}(0) - G_{\rm LR}(\gamma) \rangle$ for $\gamma N = 0.7$, as a function of the
cut-off time $t_0$. 
(These results are averaged over 20 000 different realizations.)
The standard case (without truncation) corresponds
to the limit $t_0 \to \infty$. Upon increasing $N$,
the figure shows a systematic decrease of $\delta G_{\rm LR}$ as
well as a shift of the onset of weak localization to larger 
truncation times. 

\begin{figure}
\epsfxsize=0.9\hsize
\epsffile{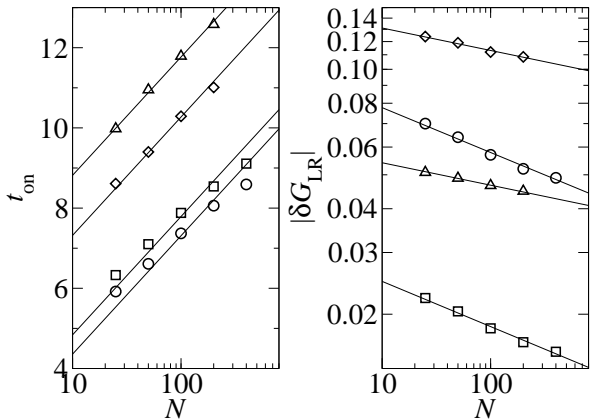}
\caption{Left: Onset time $t_{\rm on}$ obtained from the time-resolved
difference $G_{\rm LR}(\gamma) - G_{\rm LR}(0)$.
Right: Weak
localization correction $\delta G_{\rm LR}$, together with a
fit $\propto \exp(-\tau_{\rm E}/\tau_{\rm D})$. The stochasticity 
parameter $K$ uniformly chosen between $10$ and $11.5$; the 
dwell times $\tau_{\rm D}$ is set at $\tau_{\rm D} = 5$ and $\tau_{\rm
  D} = 10$. For $\tau_{\rm D} = 5$, data are shown for $\gamma N =
0.7$ (circles), and $\gamma N = 0.35$ (squares). For $\tau_{\rm D} = 10$, data are shown
for $\gamma N = 0.3$ (diamonds) and $\gamma N = 0.15$ (triangles).
\label{fig:wl2b}}
\end{figure}

\begin{figure}
\epsfxsize=0.9\hsize
\epsffile{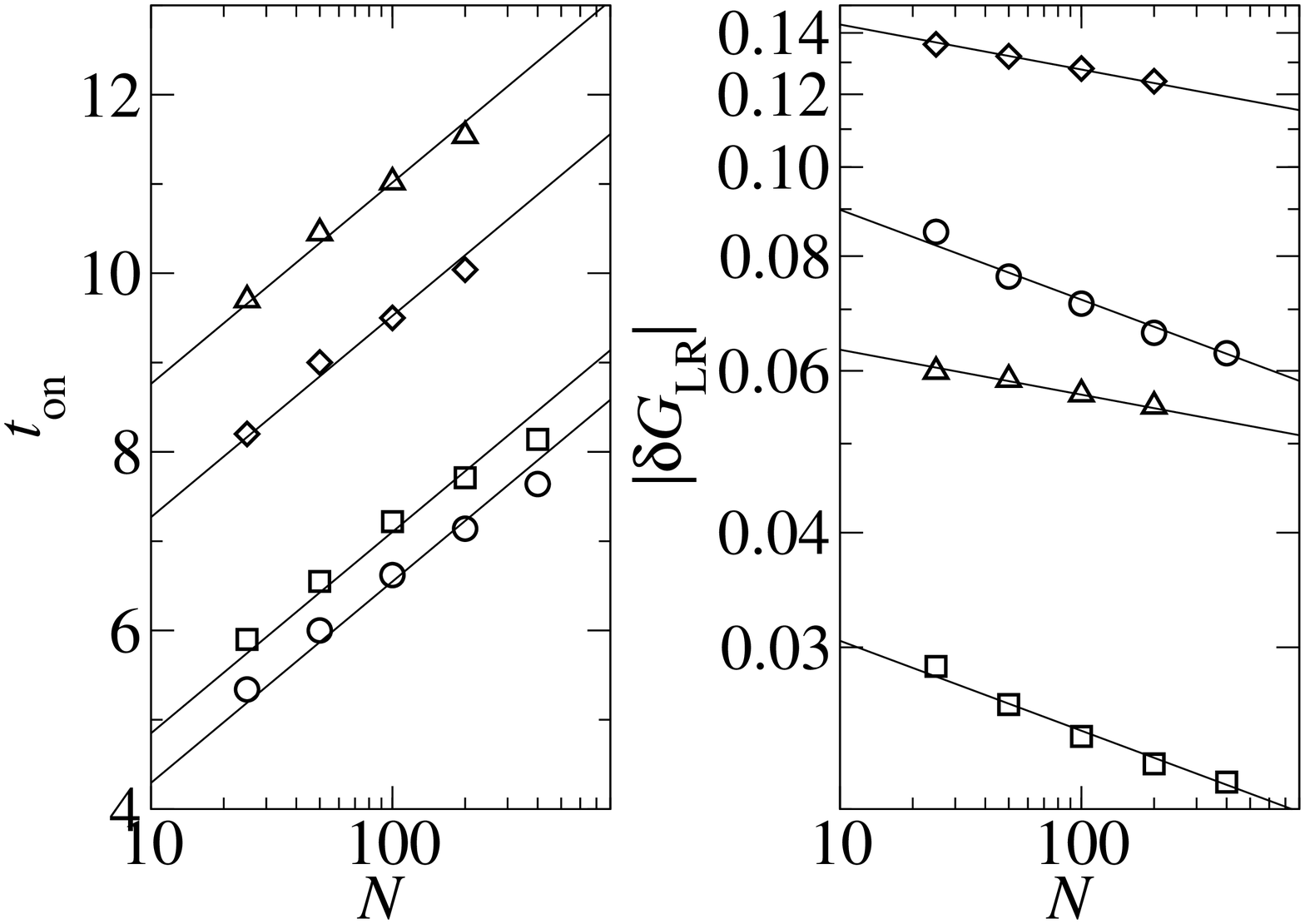}
\caption{Same as Figure {\protect{\ref{fig:wl2b}}}, but for $K$ randomly chosen between $20$ and $23$.\label{fig:wl2c}}
\end{figure}

To study these numerical result quantitatively, we examine the
dependence of the onset of weak localization and its magnitude
on the number of channels.
As an operational definition of the 
onset time $t_{\rm on}$, we define $t_{\rm on}$ as that 
truncation time $t_0$
for which $\delta G_{\rm LR} (t_0) = 
0.1 \delta G_{\rm LR}(t_0 \to \infty)$. 
In principle, onset times could also have been defined using $\delta
G_{\rm LL}$. We prefer to use $\delta G_{\rm LR}$, since the latter
are less impacted by the discreteness of the map's time evolution.
The left panel of Fig.\ \ref{fig:wl2b} shows $t_{\rm on}$ as a 
function of $N$ for the simulation curves shown in
Fig.\ \ref{fig:wl2}, as well as
for similar curves calculated for $\gamma N = 0.3$ (time-resolved
data not shown) and for simulation data taken at
dwell time $\tau_{\rm D} = 10$.
The right panel of Fig.\ \ref{fig:wl2b} shows the
dependence of $|\delta G_{\rm LR}|$ on $N$. 
Figure \ref{fig:wl2c} depicts the same data as figure
\ref{fig:wl2b}, but for stochasticity parameter $K$ taken uniformly
in the interval $20 < K < 23$.

According
to the semiclassical theory, the derivative $d t_{\rm on}/d \ln N
= 2/\lambda$, where $\lambda$ is the Lyapunov exponent
of the map.
Similarly, semiclassics predicts that
$d \ln |\delta G_{\rm LR}|/d \ln N = -1/\lambda \tau_{\rm
  D}$. The classical map 
corresponding to Eq.\ (\ref{3kick}) is described in Ref.\
\onlinecite{kn:tworzydlo2004b}, and its Lyapunov exponent is readily
calculated using the method described in Ref.~ \onlinecite{kn:ott2002}.
Since the magnetic fields that break time reversal invariance
are classically small, the Lyapunov exponents should be computed
for $\gamma=0$. Numerically, we find $\lambda = 1.56$ for 
$K \approx 10$ and $\lambda = 2.04$ for $K \approx 20$. Lines
with slopes corresponding these Lyapunov exponents are shown in Figs.\
\ref{fig:wl2b} and \ref{fig:wl2c}. We conclude that the $N$-dependence
of $| \delta G_{\rm LR}|$ is well described by the semiclassical
theory. The onset times deviate somewhat from the expected
slope, however. Since this deviation is more pronounced for the shorter dwell
time we attribute it to fluctuations in the values of the Lyapunov 
exponent. Fluctuations of the Lyapunov exponent have a stronger effect 
on quantum corrections at short dwell times than at large dwell
times.\cite{kn:aleiner1996} 

Although our data are obtained in the same way as in Ref.\
\onlinecite{kn:tworzydlo2004b}, our conclusions are markedly
different. Whereas Ref.\ \onlinecite{kn:tworzydlo2004b}
finds no evidence of a systematic Ehrenfest-time dependence of 
weak localization, we conclude that there is a systematic 
Ehrenfest-time dependence of weak localization and that
the simulation results for weak localization are consistent with
the semiclassical theory. 
We attribute the differences to a lack of accuracy in the
simulations of Ref.\ \onlinecite{kn:tworzydlo2004b}. Indeed, on
average the simulations of Ref.\ \onlinecite{kn:tworzydlo2004b} do
show a slight decrease upon increasing $N$, but the statistical
uncertainties are too large to rule whether the decrease is systematic
or accidental.

\begin{figure}
\epsfxsize=0.9\hsize
\epsffile{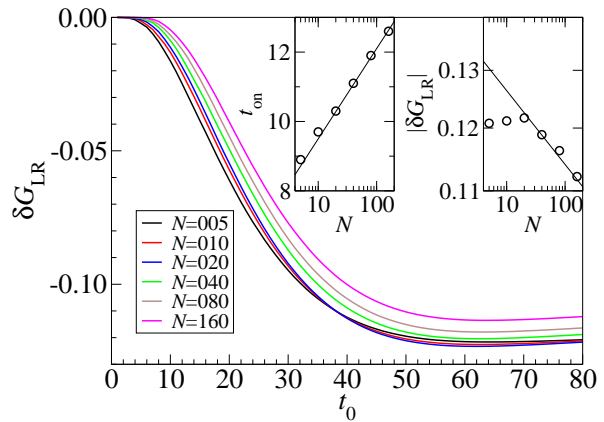}
\caption{Time-resolved 
difference between ensemble-averaged
conductance at $\gamma N = 0$ and $\gamma N = 0.2$ for the three-kick
model (\protect\ref{3kick}), for $K \approx 10$ and 
$\tau_{\rm D} = 12.8$. Left inset: onset times. Right inset:
weak localization correction.
\label{fig:wl6}}
\end{figure}

Figure \ref{fig:wl6} shows time-resolved data for $\delta G_{\rm LR}$
for $\tau_{\rm D} = 12.8$, $q = 0.3$, $\gamma N = 0.2$, and $K$
taken uniformly in the interval $10 < K < 11.5$. Whereas the
simulation data are consistent with the semiclassical 
theory for $N \gtrsim 20$ (see the insets in Fig.\ \ref{fig:wl6}),
the simulation results for $\delta G_{\rm LR}$ show significant 
deviations from the semiclassical theory for smaller channel 
numbers. This is no surprise, as the semiclassical theory is known
to break down for small $N$. Despite the differences between the
semiclassical theory and the magnitude of the simulated weak
localization correction $\delta G_{\rm LR}$ for small $N$, the onset times 
appear to
depend linearly on $\ln N$ down to the smallest channel numbers
considered in the simulation ($N=5$).


\begin{figure}
\epsfxsize=0.9\hsize
\epsffile{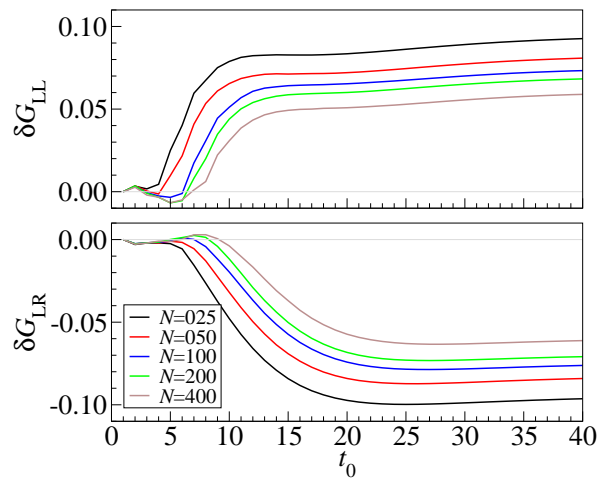}
\caption{\label{fig:wl3} Weak localization correction to reflection
and transmission coefficients, obtained using Eq.\
(\protect\ref{eq:scaling}).
Data shown are for $K \approx 10$ and $\tau_{\rm D} = 5$. The
average was taken over an ensemble of $80\, 000$ realizations, except for $N=400$, as described in the text.}
\end{figure}

\begin{figure}
\epsfxsize=0.9\hsize
\epsffile{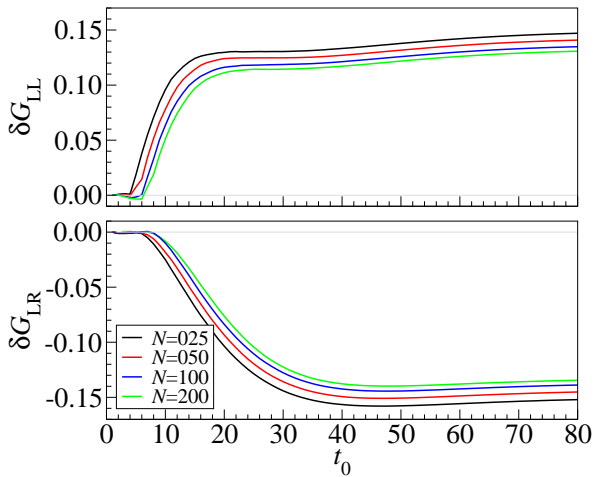}

\caption{\label{fig:wl4} Weak localization correction to reflection
and transmission coefficients, obtained using Eq.\
(\protect\ref{eq:scaling}).
Data shown are for $K \approx 10$ and $\tau_{\rm D} = 10$. The
number of realizations used to calculate the ensemble average is
$20\, 000$.}
\end{figure}

\begin{figure}
\epsfxsize=0.9\hsize
\epsffile{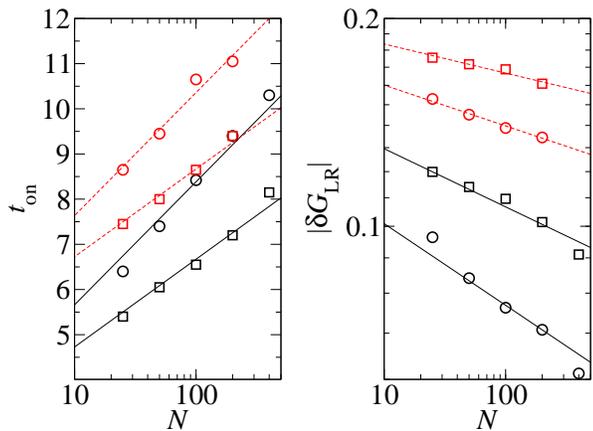}
\caption{Left: Onset time $t_{\rm on}$ obtained from the time-resolved
weak localization correction $\delta G_{\rm LR}$.
Right: Weak
localization correction $\delta G_{\rm LR}$, together with a
fit $\propto \exp(-\tau_{\rm E}/\tau_{\rm D})$.  Data are taken with
stochasticity parameter $K$ uniformly chosen between $10$ and $11.5$
(circles) and between $20$ and $23$ (squares), and for dwell times
$\tau_{\rm D} = 5$ (solid) and $\tau_{\rm D} = 10$ (dashed). 
\label{fig:wl4b}}
\end{figure}

The second method of computing weak localization uses the scaling 
of the conductance with the number of channels, see Eq.\
(\ref{eq:scaling}). Time-resolved results for reflection and
transmission coefficients are shown in Fig.\ \ref{fig:wl3} and
\ref{fig:wl4}, with $K$ taken randomly from the interval
$10 < K < 11.5$ and dwell times $\tau_{\rm D} = 5$ and $\tau_{\rm D}
= 10$, respectively. For $\tau_{\rm D} = 5$, an average was taken
over $80\, 000$ samples, except for the data point at $N=400$.
Due to the numerical cost of the calculations for $N=400$ only
$40 \, 000$ samples were considered for $K\approx 10$ and 
$20 \, 000$ samples for $K \approx 20$.
For $\tau_{\rm D} = 10$, the average was
taken over $20\, 000$ samples for $K \approx 10$ and over 
$40\, 000$ samples for $K \approx 20$. 
In all cases the statistical error
of the weak localization correction was estimated to be $\sim
10^{-3}$. Again, we compare the data in Fig.\ \ref{fig:wl4b} to 
lines with slopes which are fixed by the Lyapunov exponent of the 
corresponding classical map. For this model the Lyapunov exponents 
are $1.69$ when $K \approx 10$ and $2.37$ when $K\approx 20$.
As for the
three-kick model, the results are consistent with the semiclassical
theory, see the right panel of Fig.\ \ref{fig:wl4b}. Figure 
\ref{fig:wl4b} also contains results for $K$ taken uniformly in
the interval $20 < K < 23$ (time-resolved data not shown).

The time-resolved data shown in Figs.\ \ref{fig:wl3} and \ref{fig:wl4}
show some structure at times smaller than the onset time. We cannot
rule out that this structure, which appears to persist despite 
statistical averaging, affects our quantitative conclusions for the onset
times and weak localization corrections. Since there is no such 
small-time structure in the weak localization data obtained by
varying the field $\gamma$, we conclude that the small-time feature in the
data of Figs.\ \ref{fig:wl3} and \ref{fig:wl4} must be a
non-magnetic-field-dependent quantum correction ({\em e.g.}, 
resulting from diffraction at the contacts). The feature disappears
quickly upon increasing the stochasticity parameter $K$.

The largest channel numbers in our simulations --- $N=200$ and
$N=400$ for the scaling method, with an average over $80\, 000$
and $40\, 000$ samples (for $K \approx 10$), 
respectively --- are smaller than the 
largest channel number used in the
simulations Ref.\ \onlinecite{kn:tworzydlo2004b}. The observed
decrease of the weak localization correction is statistically
significant and systematic, but small. We could
not obtain sufficiently accurate simulation data for higher number of channels.
Increasing $N$ at fixed numerical cost implies that the
number of realizations in the average has to scale $\propto N^{-2}$,
so that the statistical error scales $\propto N$. Obtaining
simulation data for $N=800$ at the same numerical cost as our $N =
200$ data, would mean that only $5000$ samples can be averaged
over, increasing the error by a factor $4$. At that point the
statistical error becomes comparable to the expected incremental 
decrease of the weak localization data, and the simulation data loose 
statistical significance.

%
%

\begin{figure}
\epsfxsize=0.9\hsize
\epsffile{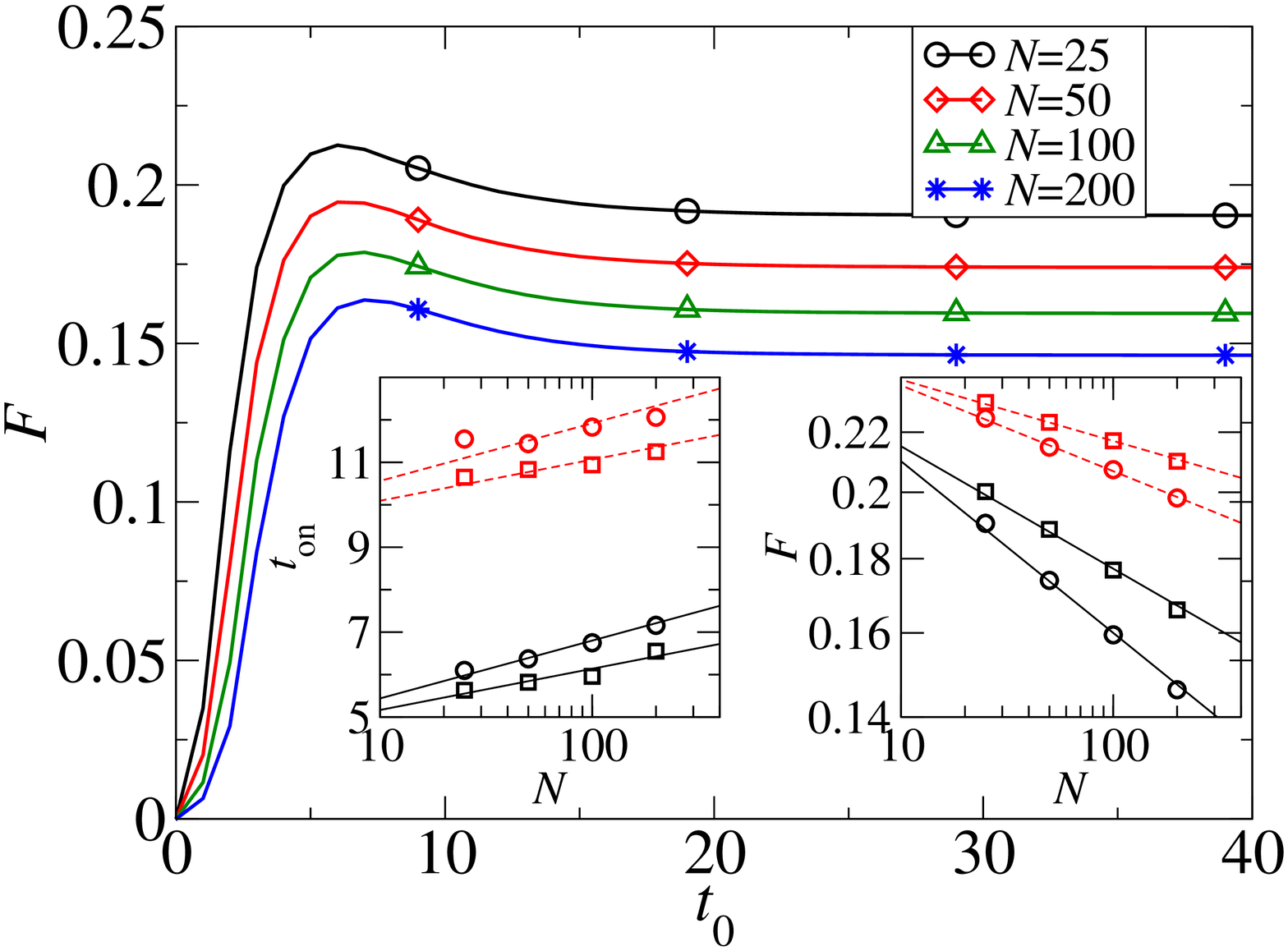}
\caption{Time-resolved shot noise Fano factors for the open kicked
rotator, with $K = 10$ and $\tau_{\rm D} = 5$. Left inset: Onset
times, determined from the location of the maximum of time-resolved
Fano factor as a function of the truncation time $t_0$. 
Right inset: Fano factor compared to a fit $\propto 
\exp(-\tau_{\rm E}/\tau_{\rm D})$.  Results for $K \approx 10$ (circles),
$K \approx 20$ (squares), $\tau_{\rm D}=5$ (solid lines), and 
$\tau_{\rm D}=10$ (dashed lines) are depicted in the insets.
\label{fig:sn1}}
\end{figure}

{\em Shot noise.} Figure \ref{fig:sn1} shows time-resolved data
for the shot noise Fano factor $F$, taken for the one-kick model
with stochasticity parameter $K \approx 10$ and dwell time $\tau_{\rm D} 
= 5$. The statistical average
was taken over $20$ realizations only, which is sufficient as the
Fano factor is self-averaging for large channel numbers. 
A more quantitative analysis of the data for $K \approx 10$ (circles)
, $K \approx 20$ (squares), $\tau_{\rm D}=5$ (solid lines), and
$\tau_{\rm D}=10$ (dashed lines) is presented in the insets.
Onset
times, determined from the maximum of the $F(t_0)$ graphs, are
shown in the left inset of Fig.\ \ref{fig:sn1}, together with a
linear fit $\propto \tau_{\rm E} + \mbox{const}.$, using the Lyapunov
exponents obtained from the corresponding classical map. The time of the
maximum was defined as onset time since it is
problematic to obtain reliable onset times using the definition
used previously. However, numerically obtained onset times for shot
noise are significantly impacted by the discreteness of the map's time
evolution, and do not show a smooth linear increase with $N$ over the
range of channel numbers investigated in our simulations.
The right
inset of the figure shows values of the Fano factors for the limit of large
truncation times, together with fits $\propto \exp(-\tau_{\rm
  E}/\tau_{\rm D})$, with the Lyapunov exponents $\lambda$ 
obtained from the corresponding classical map. 
We conclude that the rate of exponential suppression of
shot noise upon increasing $N$ is consistent with the rate of 
exponential suppression of weak localization.

\begin{figure}[thb]
\epsfxsize=0.85\hsize
\epsffile{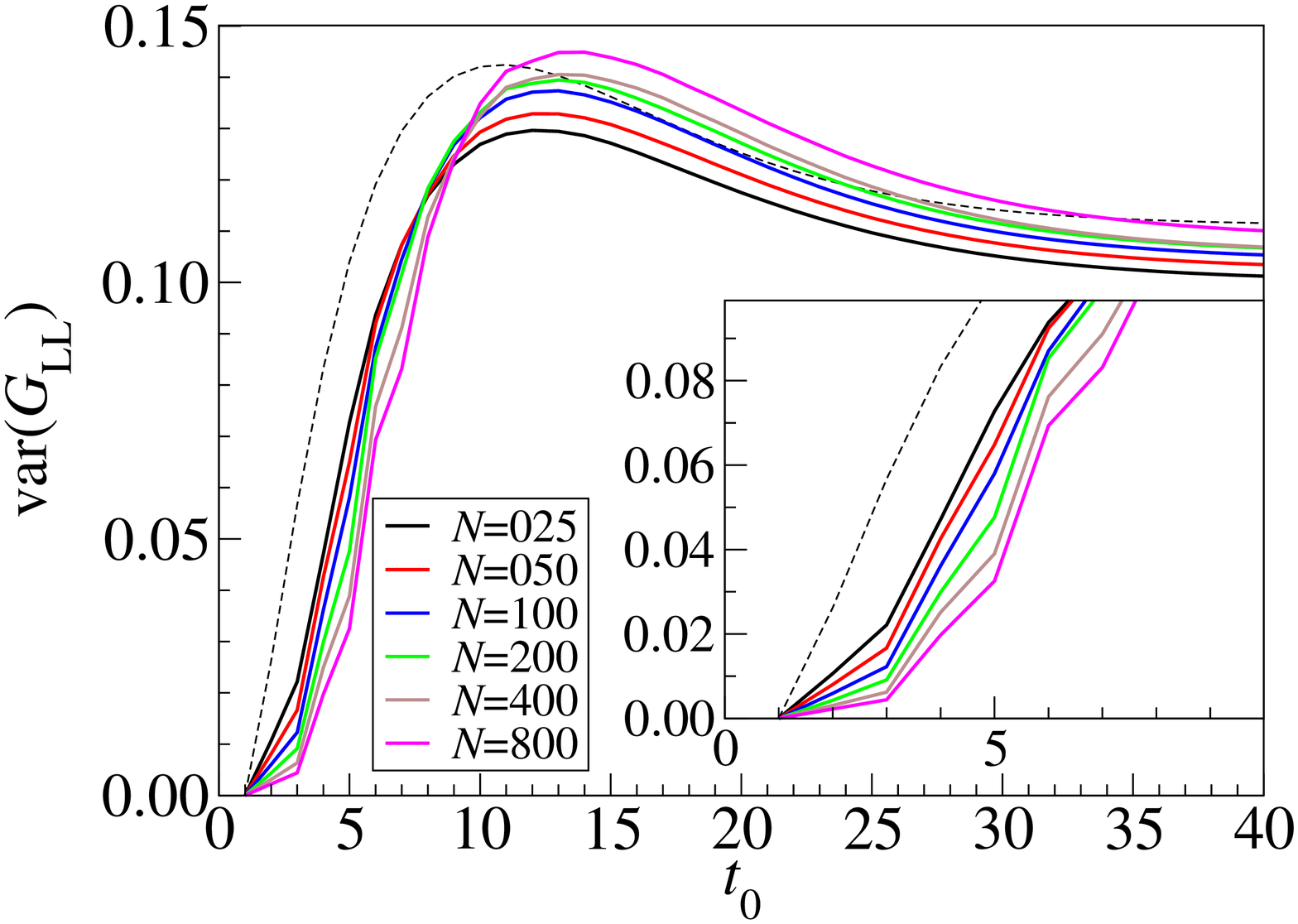}\vspace{0.5cm}

\epsfxsize=0.85\hsize
\epsffile{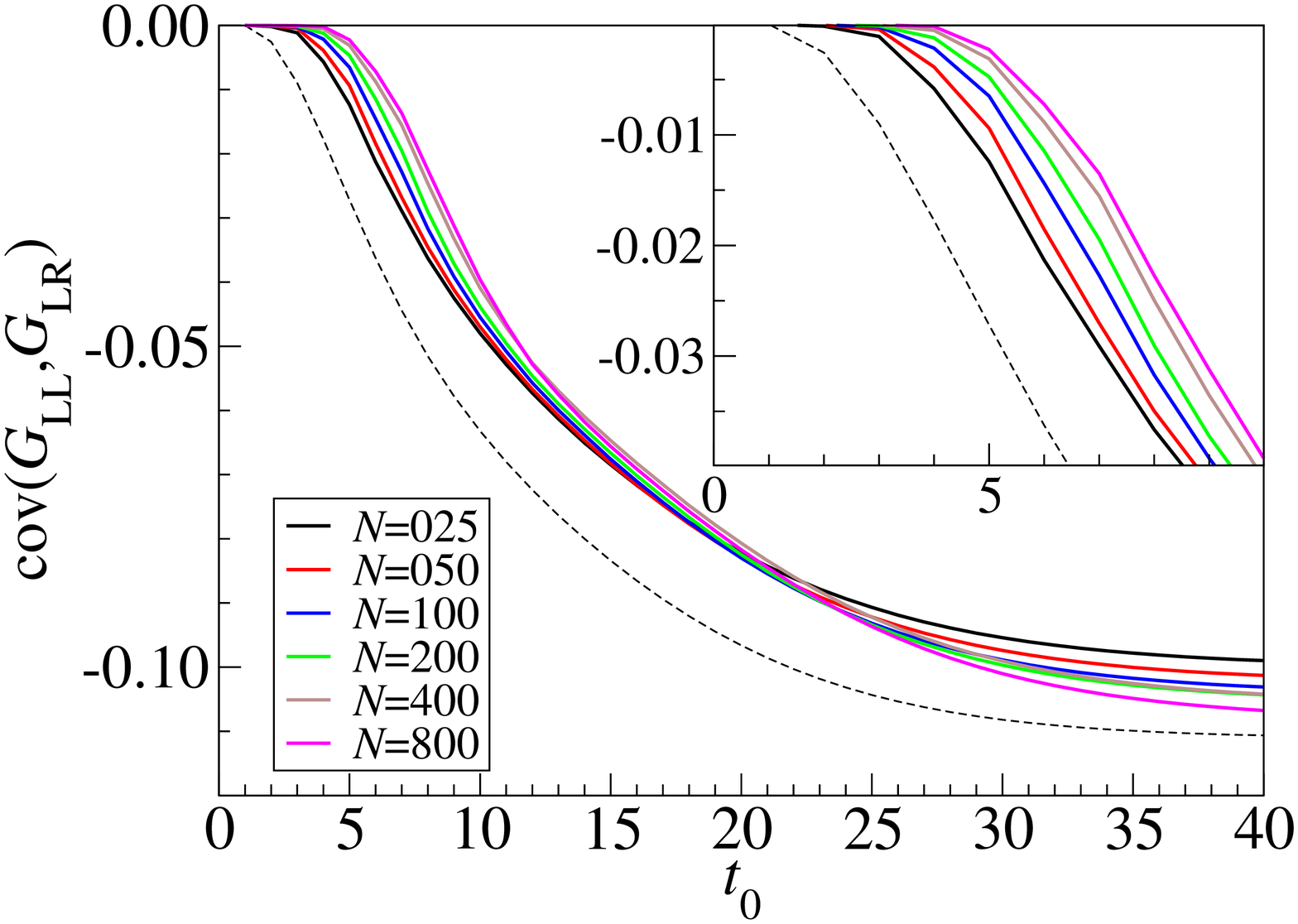}\vspace{0.5cm}

\epsfxsize=0.85\hsize
\epsffile{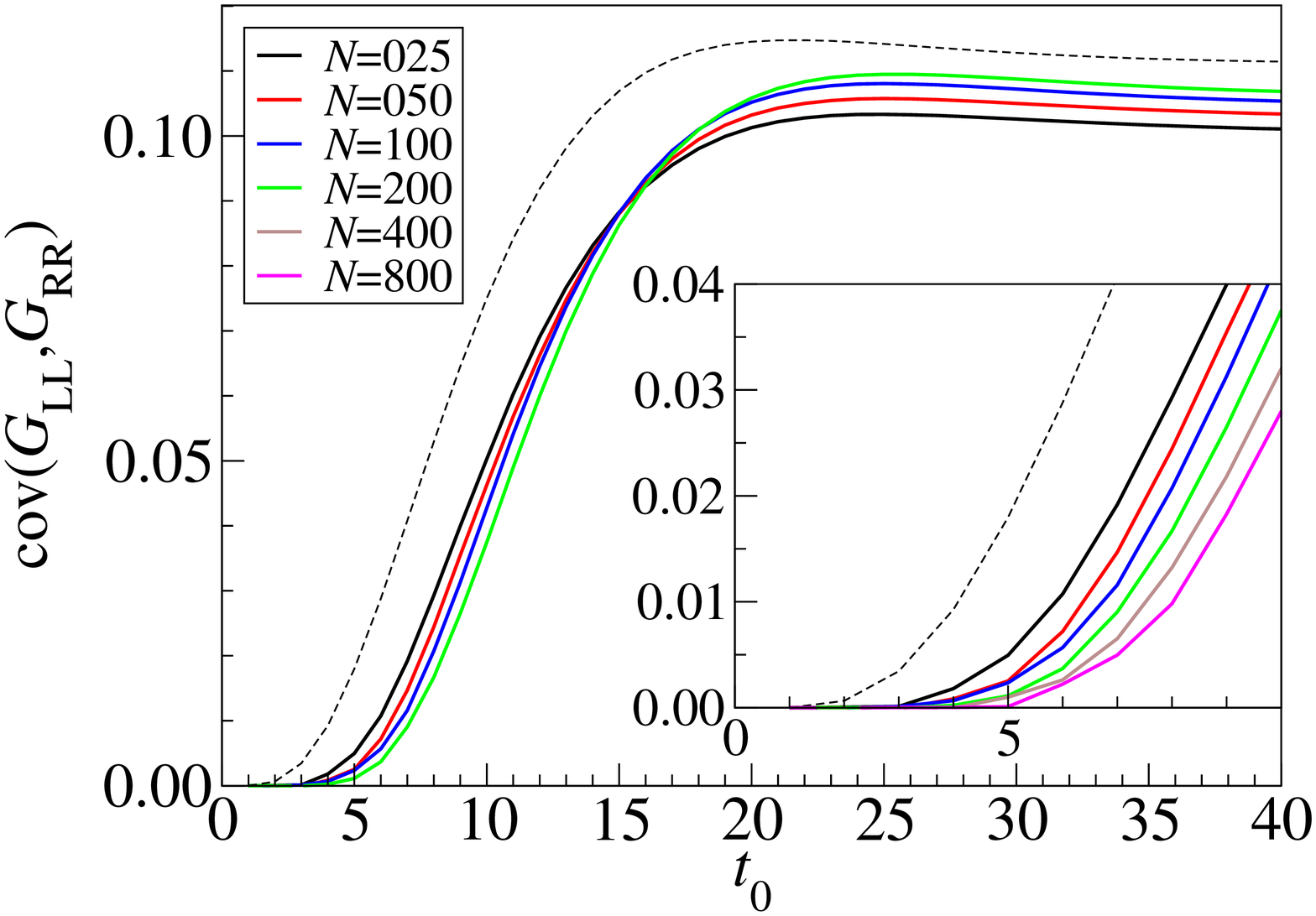}

\caption{The truncation time-dependence of $\mbox{var}\, G_{\rm LL}$,
  $\mbox{cov}\, (G_{\rm LL},G_{\rm LR})$, and $\mbox{cov}\, (G_{\rm
  LL}, G_{\rm RR})$ for the one-kick rotator. The dwell time is
  $\tau_{\rm D}=5$, while $K$ is chosen uniformly between $10$ and
  $11.5$. The corresponding time dependence given by random matrix
theory is shown dashed. The short-time behavior is magnified in the 
inset. \label{fig:ucf1}}
\end{figure} 

%
%


\begin{figure}
\epsfxsize=0.9\hsize
\epsffile{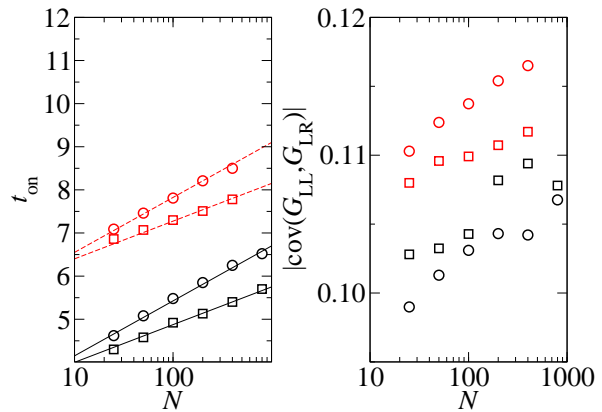}
\caption{Left: Onset time $t_{\rm on}$ obtained from the time-resolved
conductance covariance data.
Right: Conductance covariance $\cov(G_{\rm LR},G_{\rm LL})$. 
Data are taken with
stochasticity parameter $K$ uniformly chosen between $10$ and $11.5$
(squares) and between $20$ and $23$ (circles), and for dwell times
$\tau_{\rm D} = 5$ (solid) and $\tau_{\rm D} = 10$ (dashed).
\label{fig:ucf3}}
\end{figure}

{\em Conductance fluctuations.} We performed
time-resolved simulations for the variance of the reflection
from the left contact $G_{\rm LL}$, the covariance of reflection 
$G_{\rm LL}$ and transmission $G_{\rm LR}$, and the covariance of 
reflections from the left and right contacts.
We have computed the variance of the conductance coefficient 
$G_{\rm LL}$, as 
well as the covariances $\mbox{cov}\, (G_{\rm LL},G_{\rm LR})$ and
$\mbox{cov}\, (G_{\rm LL},G_{\rm RR})$ as a function of the truncation
time $t_0$. Unitarity of the scattering matrix implies that
$\mbox{var}\, G_{\rm LL} = - \mbox{cov}\, (G_{\rm LL},G_{\rm LR}) =
\mbox{cov}\, (G_{\rm LL},G_{\rm RR})$ if $t_0 \to \infty$, but
this equality does not hold for finite $t_0$.
Our simulations were done for dwell
times $\tau_{\rm D} = 5$ and $10$, and for stochasticity 
parameters $K$ between $10$ and $11.5$ and between $20$ and $23$.
Results for $\tau_{\rm D} = 5$ and
$K \approx 10$, which are representative for all results 
obtained, are shown in Fig.\ \ref{fig:ucf1}.
The number of realizations used for the averages shown here is
$80\, 000$ for $\tau_{\rm D} = 5$, except for the $N=800$ data
point, for which only $40\, 000$ realizations were taken. 
The statistical error of the variances was estimated to be 
$\sim 10^{-4}$. Onset times calculated from
the time-resolved covariance of $G_{\rm LL}$ and $G_{\rm LR}$ are shown
in Fig.\ \ref{fig:ucf3}, together with conductance covariances.
The scatter of the data points in the right panel of Fig.\
\ref{fig:ucf3} for the short dwell time $\tau_{\rm D} = 5$ is larger
than our statistical error and reproducable. It is probably an
artefact of the discrete time evolution of the map. (Note that similar
scatter exists for the weak localization data, although, in that case,
it is obscured by the large systematic decrease of the weak
localization correction upon increasing $N$.)

As can be seen from the left panel of Fig.\ \ref{fig:ucf3}, the 
onset time depend linearly
on $\ln N$: $\tau_{\rm on} \approx 0.55 \ln N + \mbox{const}.$ 
for $K \approx 10$ and
$\tau_{\rm on} \approx 0.38 \ln N + \mbox{const.}$ for $K \approx 
20$. However, the slopes are a factor $2.2$ smaller than 
for the weak localization data, both for $K \approx 10$ and $K \approx
20$. Further, the conductance variance ({\em i.e.,}
the large-$t_0$ limit of the data shown in Fig.\ \ref{fig:ucf1}) 
shows a slight {\em increase} with increasing 
$N$. The total increase is less than $10 \%$ over the range of
channel numbers considered in our simulations and falls within the
statistical uncertainty of simulations reported in Refs.\
\onlinecite{kn:jacquod2004,kn:tworzydlo2004,kn:tworzydlo2004c}, where
it was concluded that $\mbox{var}\, G$ is independent of $N$. 
Simulations of conductance fluctuations for the three-kick model
slow a less than 10\% decrease of $\mbox{var}\, G$ for the same range of
$N$, as well as onset times that increase significantly slower
with $N$ than the onset times for weak localization in the three-kick
model (data not shown).

Clearly, the simulation data for conductance
fluctuations are qualitatively different from the simulation data
for weak localization; they differ both with respect to the
$N$-dependence of the onset times and the $N$-dependence of the
magnitude of the quantum corrections. 
This contradicts the notion that the same interference processes
(diffusons and cooperons and their generalizations to ballistic
systems) underly both weak localization and conductance fluctuations,
although one should note that
there is no semiclassical
theory of the Ehrenfest-time dependence of
conductance fluctuations yet.


\section{Quantum transport through chaotic cavities: semiclassical
  theory}
\label{ALcalculation}

A semiclassical theory for the Ehrenfest-time dependence of the
weak localization correction in chaotic cavities was first 
formulated by Aleiner and Larkin.\cite{kn:aleiner1996} That theory
predicts that the weak localization correction is suppressed
$\propto \exp(-2 \tau_{\rm E}/\tau_{\rm D})$. However, the analysis of
Ref.\ \onlinecite{kn:aleiner1996} does not account for all
classical correlations: it neglects the notion that no quantum 
diffraction takes place for electrons that spend less than a 
time $\tau_{\rm E}$ in the cavity. In this section we show that
accounting for all classical correlations gives a weak
localization correction that still depends exponentially on 
the Ehrenfest time, but with a different exponent: $\delta G
\propto \exp(-\tau_{\rm E}/\tau_{\rm D})$. While a part of
our calculations has already appeared elsewhere,\cite{kn:rahav2005}
this section includes a calculation of the full conductance
matrix $G_{\alpha \beta}$. This allows to verify that unitarity 
is preserved. As the possible loss of unitarity is known to be a
problem  in semiclassical theories, it is important to
demonstrate explicitly that the method used here preserves
probability.
 
\begin{figure}[htb]
\includegraphics[width=\columnwidth]{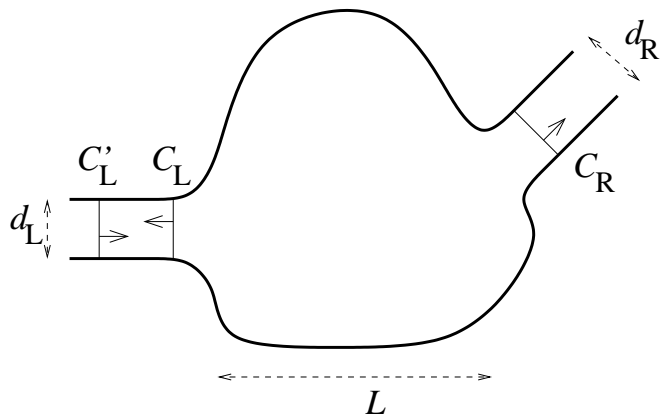}
\caption{Schematic picture of the ballistic chaotic cavity of size
$L$. The cavity is attached to two leads, labeled ``L'' and ``R'',
and with width $d_{\rm L}$ and $d_{\rm R}$, respectively. The 
contours $C_{\rm L}'$, $C_{\rm L}$, and $C_{\rm R}$ drawn in the 
leads are used for the calculation of transmission and reflection
coefficients, see the text.
\label{dot1}} 
\end{figure}

The system under consideration is a
ballistic cavity with chaotic classical dynamics. The cavity is
two-dimensional, and it is coupled to two electron reservoirs via 
contacts of width $d_{\rm L}$ and $d_{\rm R}$, see Fig.\ \ref{dot1}.
Ergodicity of the electron motion inside the dot is ensured by
the condition $d_{\rm L}, d_{\rm R} \ll L$.
The cavity and the contacts are considered in the semiclassical 
limit $d_{\rm L} k, d_{\rm R} k \gg 1$, where $k$ is
the electron wavenumber. 
Thus, one has the separation of length scales $1/k \ll 
d_{\rm L}, d_{\rm R} \ll L$ or, equivalently, the separation
of time scales $(v k)^{-1} \ll \tau_{\rm erg} \ll \tau_{\rm
  D}$, where $v$ is the electron velocity, $\tau_{\rm erg}
\sim L/v$ the ergodic time, and $\tau_{\rm D}$ the dwell time.
The latter can be expressed as
\begin{equation}
\label{taud}
\tau_{\rm D}^{-1} = \frac{(d_{\rm L} + d_{\rm R}) v}{\pi {\cal A},}
\end{equation}
where ${\cal A}$ is the area of the cavity.
We also define
\begin{equation}
  P_{\rm R} \equiv \frac{d_{\rm R}}{d_{\rm R} + d_{\rm L}},
 \ \ \
  P_{\rm L} \equiv \frac{d_{\rm L}}{d_{\rm R} + d_{\rm L}}
  \label{eq:PLR}
\end{equation}
as the probabilities that an electron at a random point in the
cavity will leave through the right and left contacts, respectively.
Electron motion inside the
cavity is assumed to be fully phase coherent and electron-electron 
interactions are neglected.

The classical motion in the cavity is assumed to be fully
chaotic (when closed), with a Lyapunov exponent $\lambda$. This exponent
measures the rate of divergence of close phase space points. Since this
divergence is due to scattering from the curved walls of the cavity one 
can expect that $\lambda \sim 1/\tau_{\rm erg}$. 
We neglect fluctuations in the Lyapunov exponent and
assume that its value is uniform. This is done to simplify the calculation
as much as possible and should not affect the main physical
conclusions as long as relevant times are much longer than $\tau_{\rm
  erg}$. 


The Aleiner-Larkin formalism is based on a semiclassical description of 
transport. Central objects in their theory are the non-oscillating
parts of the product of an advanced and retarded Green functions.
The ``diffuson'' is composed of combinations of orbits with
themselves (known as the diagonal approximation in the semiclassical 
community) while the ``cooperon'' is composed of orbits and their time
reversal counterparts.
The equation of motion for the diffuson, neglecting quantum phenomena,
is the Liouville equation,
\begin{equation}
\label{diffusoncl}
\left[ -i \omega + \hat{L}_1 \right] 
  {\cal D}^0 ( \omega ; 1 ,2 ) = \delta(1,2),
\end{equation}
where $j\equiv (\phi_j, {\bf R}_j)=1,2$ denotes the phase space coordinates,
limited to the energy shell: ${\bf R}$ is the electron 
position while the angle $\phi$ represents is the direction of the electron's
velocity $\vv$. The operator $\hat{L}_1$ is the Liouville operator,
taken with respect to the coordinates $(\phi_1, \vR_1)$.
For a hard wall cavity, $\hat{L}= \vv \cdot {\partial}/{\partial
  \vR}$, with appropriate boundary conditions at the walls.
The symbol $\delta(1,2)$ denotes a delta function on the energy shell, that is,
$\delta(1,2) \equiv 2 \pi
\delta ( {\bf R}_1-{\bf R}_2) \delta ( \phi_1-\phi_2)$. 

Quantum effects are introduced by including a weak additional 
random quantum potential, treating it as a perturbation, and averaging
over its realizations.\cite{kn:aleiner1996}
This random potential leads to two different effects, which are expected in a 
quantum theory. The first is a diffusion in phase space, which mimics
the fact that quantum dynamics cannot separate
small phase space structures. 
The second effect is 'quantum switching' between close
phase space trajectories, which results in weak localization.
The strength $\gamma_{\rm q}$
of these effects is then set so that quantum corrections
will appear after the correct time, the Ehrenfest time $\tau_{\rm E}$.
Phase space diffusion modifies the evolution equation for the 
leading order diffuson as\cite{kn:aleiner1996}
\begin{equation}
\label{diffusonw}
\left[ -i \omega + \hat{L}_1 - \gamma_{\rm q} \frac{\partial^2}{\partial \phi_1^2} \right] {\cal D}^0 ( \omega ; 1 ,2 ) = \delta(1,2).
\end{equation}
the strength $\gamma_{\rm q}$ of the phase space diffusion term
is estimated to be~\cite{kn:aleiner1996} $\gamma_{\rm q}
\simeq \lambda^2/k v$.
It is instructive to note that this term, although small, 
breaks the time reversal invariance which is a property of {\em both} 
quantum mechanics and classical mechanics (in absence of a 
magnetic field). In the presence of quantum diffraction, 
the leading order cooperon, ${\cal C}^0$, also satisfies
Eq.\ (\ref{diffusonw}).

In addition to the presence of the phase space diffusion term in the
evolution equation (\ref{diffusonw}), quantum diffraction leads to
an quantum correction to the diffuson: ${\cal D} = {\cal D}^0 + 
\Delta {\cal D}$. This quantum correction is referred to as 
weak localization, because similar quantum corrections are the
precursor of Anderson localization in disordered systems. 
The weak localization correction for the diffuson 
reads\cite{kn:aleiner1996}
\begin{widetext}
\begin{multline}
\label{dd}
\Delta {\cal D} (\omega; 1 , 2) =  {\cal D}^0 (\omega;  1 , \bar{2}) 
\frac{{\cal
    C}^0 (\omega; \bar{2} , 2)}{2 \pi \hbar \nu} +  \frac{{\cal C}^0 (\omega; 1,
  \bar{1})}{2 \pi \hbar \nu} {\cal D}^0 (\omega; \bar{1}, 2 ) 
 \\
 +  \int d3 {\cal D}^0 (\omega; 1,3) {\cal D}^0 (\omega; \bar{3},2) 
 \left[ 2 i \omega - \hat{L}_3 + \gamma_{\rm q} \frac{\partial^2}{\partial \phi_3^2} \right] \frac{{\cal C}^0 (\omega;  3,\bar{3})}{2 \pi \hbar \nu},
\end{multline}
where $\bar{j} = (\phi_{j}+\pi, {\bf R}_j)$ denotes the time reversal of 
phase space point $j$ and 
$\nu={m}/{2 \pi \hbar^2}$ is the density of states per unit area.
In terms of classical trajectories, the weak localization correction 
$\Delta {\cal D}$ corresponds to a configuration as shown in Fig.\
\ref{fig:1}a: $\Delta {\cal D}$ is the correction to the product of
a retarded and advanced Green function arising from the combination 
of a trajectory that intersects itself at a small angle and a
trajectory that avoids intersecting
itself.\cite{kn:aleiner1996,kn:sieber2001} If the intersection
angle is sufficiently small, such a pair of
trajectories has strongly correlated actions, resulting in a net
quantum correction to the diffuson even after ensemble averaging.
The pairs of trajectories before and after the `interference region'
are represented by the leading-order diffusons ${\cal D}^0$ 
in Eq.\ (\ref{dd}),
whereas the trajectories in the closed loop --- an orbit conjugate
to its time-reversed counterpart --- are represented by the
leading order cooperon ${\cal C}^0$. 

In order to calculate the weak localization correction to the
conductance, the dot's conductance coefficients $G_{\rm LL}$ and $G_{\rm
RL}$ are expressed in terms of the diffuson ${\cal D}$,
\begin{equation}
\label{totT}
G_{\rm RL} = 2 \pi \hbar \nu v^2 \int_{C_{\rm L}'} dl_1 
  \int_{\phi_{\rm L}'-\pi/2}^{\phi_{\rm L}'+\pi/2} \frac{d \phi_1}{2 \pi}
  \cos(\phi_{\rm L} - \phi_1)
  \int_{C_{\rm R}} dl_2
  \int_{\phi_{\rm R}-\pi/2}^{\phi_{\rm R}+\pi/2} \frac{d \phi_2}{2 \pi}
  \cos(\phi_{{\rm R}} - \phi_2)
  {\cal D} (2,1),
\end{equation}
and
\begin{equation}
\label{totR}
G_{\rm LL} = 2 \pi \hbar \nu v \int_{C_{\rm L}'} dl_1
  \int_{\phi_{\rm L}'-\pi/2}^{\phi_{\rm L}'+\pi/2} \frac{d \phi_1}{2 \pi}
  \cos(\phi_{{\rm L}}' - \phi_1)
  \left[
  v \int_{C_{\rm L}} dl_2 
  \int_{\phi_{\rm L}-\pi/2}^{\phi_{\rm L}+\pi/2} \frac{d \phi_2}{2 \pi}
  \cos(\phi_{\rm L} - \phi_2) 
  {\cal D} (2,1) - 1 \right].
\end{equation}
Here $C_{\rm L}$, $C_{\rm R}$, and $C_{\rm L}'$ are smooth contours
following a cross section of the contacts, see Fig.\ \ref{dot1}. The
contour $C_{\rm L}'$ is taken a small amount further away from the
cavity than the contour $C_{\rm L}$, see Fig.\
\ref{dot1}. The angles $\phi_{\rm L}$ and $\phi_{\rm R}$ denote
the direction of the outward-pointing normal to $C_{\rm L}$, $C_{\rm
  R}$, whereas the angle $\phi_{\rm L}'$ represents the direction
of the inward-pointing normal to $C_{\rm L}'$. Since we consider
DC transport, the frequency $\omega$ has been set to zero.
We refer to the Appendix for a
derivation. Similar expressions, but with an ambiguity in the precise
definitions of the integrals over the lead cross sections, were 
derived by Takane and Nakamura.\cite{kn:takane1997} 

The leading order ``classical'' transmission and reflection
coefficients are found by substituting ${\cal D}^0$ for ${\cal D}$.
Assuming that dynamics in the cavity is ergodic for all relevant
time scales, one finds\cite{kn:aleiner1996,kn:takane1997}
\begin{equation}
  G_{\rm RL}^0 = - G_{\rm LL}^0 =
  \frac{2 \pi \hbar \nu {\cal A}}{\tau_{\rm D}}
  P^{\vphantom{2}}_{\rm L} P^{\vphantom{2}}_{\rm R},
\end{equation}
where the escape probabilities to the left and right leads were defined 
in Eq.\ (\ref{eq:PLR}) above.

The weak localization correction to the conductance coefficients
$G_{\rm RL}$ and $G_{\rm LL}$
is obtained by substituting Eq.\ (\ref{dd}), for the
weak localization correction $\Delta {\cal D}$ of the diffuson, into Eqs.\ 
(\ref{totT}) and (\ref{totR}). We first consider the weak localization
correction to $G_{\rm RL}$. The first two terms in (\ref{dd}) 
contain paths which leave the dot at the point of entry, and hence
these do not contribute to the transmission. Hence
\begin{multline}
\label{deltat}
\delta G_{\rm RL} =  - v^2 \int_{C_{\rm L}'} dl_1 
  \int_{\phi_{\rm L}'-\pi/2}^{\phi_{\rm L}'+\pi/2} \frac{d \phi_1}{2 \pi}
  \cos(\phi_{{\rm L}}' - \phi_1)
  \int_{C_{\rm R}} dl_2 
  \int_{\phi_{\rm R}-\pi/2}^{\phi_{\rm R}+\pi/2} \frac{d \phi_2}{2 \pi}
  \cos(\phi_{\rm R} - \phi_2) 
  \\ \times
  \int d3 {\cal D}^0 (2,3) 
  {\cal D}^0 (\bar{3},1) \left[
  \hat{L}_3 -\gamma_{\rm q} \frac{\partial^2}{\partial \phi_3^2}
  \right] {\cal C}^0 (3,\bar{3}).
\end{multline}
\end{widetext}
The phase space point 3 can be viewed as the center of the 
'interference region' as depicted in Fig.~\ref{fig:1}a.
The two diffusons and the cooperon head into opposite directions
in phase space, so that they will sample
different parts of phase space along the trajectory. Thus, one can 
treat them as statistically uncorrelated.\cite{kn:aleiner1996}

Both the cooperon and the product of diffusons in Eq.\ 
(\ref{deltat})
exhibit non-trivial correlations since the returning 
path is strongly correlated with the outgoing one, cf.\ 
Fig.\ \ref{fig:1}a. In order to take into account 
these correlations one should examine what is meant by the
phase space points $3$ and $\bar{3}$ in Eq. (\ref{deltat}). By
virtue of the phase-space diffusion term in the evolution 
equation for the leading order diffuson, effectively,
the points $3$ and
$\bar 3$ do not need to be exactly time reversed phase space
points, since quantum
mechanics allows for a finite phase space uncertainty.
An intuitive way to calculate the correlations between $3$ and $\bar{3}$
was shown by Vavilov and Larkin.\cite{kn:vavilov2003}
Instead of using the diffusion in phase space, Vavilov and Larkin 
average over a
range of initial phase space points near $3$ and final phase space 
points near $\bar{3}$. They show that the effect of the phase space 
diffusion is equivalent to such averaging (up to
logarithmic corrections) if the `size' 
of phase space area near $3$ and $\bar{3}$
scales as $k^{-1/2}$. [To be precise, the range of coordinates
$\vR$ in the average is $\sim (L/k)^{1/2}$, whereas the angular
range is $\sim (L k)^{-1/2}$.]
We adopt the procedure of Vavilov and Larkin
and consider an average over
phase space points $3'$ and $3''$ that are within a
distance of order $k^{-1/2}$ of the phase space point $3$. 
We can then follow the classical orbits which start at the points 
$3'$, $3''$. The phase space distance between these
 trajectories will diverge exponentially due to
the chaotic dynamics. Since the trajectories
 start from phase space points at a distance $\sim k^{-1/2}$,
it will take a time $\tau_{\rm E}/2$ to reach a phase space distance
of order $L$. Once the phase
space distance between the trajectories is large enough they can be 
considered as totally uncorrelated. At this point, ergodic dynamics
can be assumed. The loss of correlations happens on a time scale
of $1/\lambda$ around the time $\tau_{\rm E}/2$. Since $1/\lambda
\ll \tau_{\rm E}$ in the parameter regime where the Ehrenfest time
affects quantum transport, we can view this process as effectively
instantaneous. 

Let us first consider the cooperon in Eq.\ (\ref{deltat}). As a
result of the phase space correlations described above, one cannot
close the orbit from $\bar 3'$ to $3''$ for short times. In order
to make this more quantitative, we denote by $t_j$ the time
it takes for the classical orbit starting at phase space point $j$ 
to leave the cavity through one of the two openings. Then, if 
$t_{\bar{3}}<\tau_{\rm E}/2$ we find ${\cal C}^0 (3,\bar{3})={\cal
  C}^0 (3'',\bar{3}')=0$. On the other hand, if 
$t_{\bar{3}}>\tau_{\rm E}/2$ we can propagate the phase space point
$\bar{3'}$ for a time $\tau_{\rm E}/2$ and reach phase space point 
$4$. Similarly, we can propagate $3''$ backward in time and find,
toward the phase space point $5$. This leads to 
$$ 
  {\cal C}^0 (3'',\bar{3'}) \simeq \left\{ \begin{array}{ll} 0 &
  \mbox{if $t_{\bar{3}} < {\tau_{\rm E}}/{2}$},\\ {\cal C}^0 (5,4) &
  \mbox{if $t_{\bar{3}} > {\tau_{\rm E}}/{2}$}. \end{array} \right.
$$
When the phase space points $3'$ and $3''$ are averaged over, the cooperon
will have contributions from various phase space points $5,4$, which are their
distant past (or future). The points $4$ and $5$ 
can be taken to be uncorrelated and sample the phase space with
uniform probability. Since the phase space point $3$ is eventually
integrated over, there will be contributions from a sizable fraction
 of phase space. Thus, we approximate the average contribution to the 
cooperon by replacing ${\cal C}^0 (5,4)$ by its average value
$\tau_{\rm D}/{\cal A}$. This leads to
\begin{equation}
\label{cooperon}
\left< {\cal C}^0 (3,\bar{3})\right> = 
  \frac{\tau_{\rm D}}{\cal A} \theta(t_{\bar 3} - \tau_{\rm E}/2),
\end{equation}
where $\theta(x) = 1$ if $x > 0$ and $0$ otherwise.

We address the product of two diffusons in Eq.\
(\ref{deltat}) by considering the integral
\begin{multline}
\label{intt}
  {\cal I}(3) =  - v^2 \int_{C_{\rm L}'} dl_1 
  \int_{\phi_{\rm L}'-\pi/2}^{\phi_{\rm L}'+\pi/2} \frac{d \phi_1}{2 \pi}
  \cos(\phi_{\rm L}' - \phi_1) 
  \\ \times 
  \int_{C_{\rm R}} dl_2 
  \int_{\phi_{\rm R}-\pi/2}^{\phi_{\rm R}+\pi/2} \frac{d \phi_2}{2 \pi}
  \cos(\phi_{{\rm }} - \phi_2)
  {\cal D}^0 (2,3) {\cal D}^0 (\bar{3},1).
  \nonumber
\end{multline}
Again, the integration over phase space points $3$ and $\bar{3}$ is
replaced by an average over phase space points $3'$ and $3''$ within
a distance of order $(L/k)^{1/2}$ from 3. Since the diffuson connects
the phase space points $3'$ and $3''$ with points at two different
contacts, the product of diffusons must be zero if $t_3 < \tau_{\rm
  E}/2$ where, as before, $t_3$ is the time it takes for the orbit
at phase space point $3$ to leave the system. For larger $t_3$ we 
may, again, assume ergodicity, so that we find, after averaging over
$3'$ and $3''$,
$$
{\cal I}(3) = P_{\rm R} P_{\rm L} \theta(t_3 - \tau_{\rm E}/2).
$$

Combining results, we find
\begin{equation}
  \delta G_{\rm RL} = - P_{\rm R} P_{\rm L} \frac{\tau_{\rm D}}{\cal A}
  \int d3 \theta(t_3 - \tau_{\rm E}/2)  \hat L_3
  \theta(t_{\bar 3} - \tau_{\rm E}/2).
  \label{eq:deltat2}
\end{equation}
The Liouville operator in Eq.\ (\ref{eq:deltat2}) measures the rate of
flow of probability density out of the integration range of the
phase space variable $3$.\cite{kn:aleiner1996} The
boundary of the integration range is composed of the lead phase space 
points propagated {\em backward} for a time $t_{\rm E}/2$. This leaves 
only a fraction $\exp({-\frac{\tau_{\rm E}}{2\tau_{\rm D}}})$ 
of the size of the
boundary at the lead. To estimate the integral we assume that each outgoing
direction is equally likely to be in the system when propagated backward.
However, only points which also have $t_{\bar{3}}>\tau_{\rm E}/2$ will have a 
non-vanishing cooperon, leading to an 
additional factor of $\exp({-\frac{\tau_{\rm E}}{2\tau_{\rm D}}})$. 
Collecting contributions from both leads, we find
\begin{equation}
\delta G_{\rm RL} = - P_{\rm R} P_{\rm L} e^{-\tau_{\rm E}/\tau_{\rm D}}.
  \label{eq:dtresult}
\end{equation}

It is of interest to compute also the weak localization correction 
$\delta G_{\rm LL}$. This will allow to check 
that probability is conserved.
The calculation is similar to that of the $\delta G_{\rm RL}$,
with two important differences. The first difference is that
the weak localization
correction of reflection is composed from two parts. In addition
to the third term in Eq.\ (\ref{dd}), there is a contribution
from the second term in Eq. (\ref{dd}). We write these two parts
as $\delta G_{{\rm LL}}^{(1)}$ and $\delta G_{\rm LL}^{(2)}$, 
respectively, and calculate them separately.
Substitution of the third term in Eq. (\ref{dd}) into Eq. (\ref{totR})
gives
\begin{widetext}
\begin{multline}
  \delta G_{\rm LL}^{(1)} = - v^2 \int_{C_{\rm L}'} dl_1 
  \int_{\phi_{\rm L}'-\pi/2}^{\phi_{\rm L}'+\pi/2} \frac{d \phi_1}{2 \pi}
  \cos(\phi_{\rm L}' - \phi_1) 
  \int_{C_{\rm L}} dl_2  
  \int_{\phi_{\rm L}-\pi/2}^{\phi_{\rm L}+\pi/2} \frac{d \phi_2}{2 \pi}
  \cos(\phi_{{\rm L}} - \phi_2) \\ \times
  \int d3 {\cal D}^{0}(2,3) 
  {\cal D}^0 (\bar{3},1) 
  \left[ \hat{L}_3 -\frac{1}{\tau_q} \frac{\partial^2}{\partial \phi_3^2} \right] {\cal C}^0 (3,\bar{3}).
\label{deltar1}
\end{multline}
%
The product of the two diffusons can be assumed to be uncorrelated with
the cooperon. 
The calculation of the cooperon proceeds as for the transmission
calculation. However, 
the behavior of the product of the diffusons contributing to reflection  
differ from that of the diffusons contributing to transmission.
To see that, again consider the integral
$$
  {\cal I}(3) =  - v^2 \int_{C_{\rm L}'} dl_1 
  \int_{\phi_{\rm L}'-\pi/2}^{\phi_{\rm L}'+\pi/2} \frac{d \phi_1}{2 \pi}
  \cos(\phi_{\rm L}' - \phi_1) 
  \int_{C_{\rm L}} dl_2 
  \int_{\phi_{\rm L}-\pi/2}^{\phi_{\rm L}+\pi/2} \frac{d \phi_2}{2 \pi}
  \cos(\phi_{{\rm L}} - \phi_2)
  {\cal D}^0 (2,3) {\cal D}^0 (\bar{3},1).
$$
\end{widetext}
If $t_3>\tau_{\rm E}/2$ both diffusons become uncorrelated before leaving
the system and the average value of this integral is therefore $P_{\rm
  L}^2$.
However, if $t_3<\tau_{\rm E}/2$ both diffusons will exit the dot through
the {\em same} lead, and one finds ${\cal I} = P_{\rm L}$. Hence,
$$
{\cal I} = \left\{ \begin{array}{ll} P_{\rm L} & \mbox{if
    $t_{\bar{3}} < {\tau_{\rm E}}/{2}$};\\ P_{\rm L}^2 & \mbox{if $t_{\bar{3}} > {\tau_{\rm E}}/{2}$}. \end{array} \right.
$$
This is the second difference between the calculations of $\delta
G_{\rm RL}$ and $\delta G_{\rm LL}$. We then find
\begin{equation}
  \delta G_{\rm LL}^{(1)} = - P_{\rm L} e^{-\tau_{\rm E}/2 \tau_{\rm D}}
  (1 - e^{-\tau_{\rm E}/2 \tau_{\rm D}}) - P_{\rm L}^2 e^{-\tau_{\rm
  E}/\tau_{\rm D}}.
  \label{eq:dR1}
\end{equation}
The contribution $\delta G_{\rm LL}^{(2)}$ is obtained by substituting the 
second term in the right hand side 
of Eq. (\ref{dd}) into Eq. (\ref{totR}).
The lead integral over $2$ can be calculated, resulting in
\begin{eqnarray}
  \delta G_{\rm LL}^{(2)} &=& v \int_{C_{\rm L}} dl_1 
  \int_{\phi_{\rm L}-\pi/2}^{\phi_{\rm L}+\pi/2} \frac{d \phi_1}{2 \pi}
  \cos(\phi_{\rm L} - \phi_1) {\cal C}^0(1,\bar 1)
  \nonumber \\ &=&
  P_{\rm L} e^{-\tau_{\rm E}/2 \tau_{\rm D}}.
  \label{eq:dR2}
\end{eqnarray}
Combining Eqs.\ (\ref{eq:dR2}) and (\ref{eq:dR1}), one 
finds $\delta G_{\rm LL} = \delta G_{\rm LL}^{(1)} + \delta G_{\rm
  LL}^{(2)} = - \delta G_{\rm RL}$, as expected.

The result (\ref{eq:dtresult}) shows that, according to the
semiclassical theory, the weak localization correction to transmission 
is suppressed exponentially. However, the exponent we find is
different from that of Ref.\ \onlinecite{kn:aleiner1996},
where it is reported that $\delta G_{\rm RL} = -P_{\rm L} P_{\rm R}
\exp(-2 \tau_{\rm E}/ \tau_{\rm D})$. 
The reason for the difference with Ref.\
\onlinecite{kn:aleiner1996} is that our calculation 
obeys the classical correlations following from the
separation of phase space into a `classical' and
`quantum' part, corresponding to (classical)
trajectories of length smaller or
larger than $\tau_{\rm E}$, respectively. Quantum diffraction does not 
involve the `classical' part of phase space. Calculating the cooperon
and the product of diffuson propagators assuming ergodic dynamics in
the `quantum' part of phase space only increases 
the weak localization correction
$\delta G$ by a factor $\exp(\tau_{\rm E}/\tau_{\rm D})$ with 
respect to the calculation of Ref.\ \onlinecite{kn:aleiner1996}.
The reason that $\delta G$ remains exponentially small --- but with 
exponent $\exp(-\tau_{\rm E}/\tau_{\rm D})$, not $\exp(-2 \tau_{\rm
  E}/ \tau_{\rm D})$ --- is that, according
to the semiclassical theory, weak localization
requires a minimal path length of $2 \tau_{\rm E} v$. The fraction
of `quantum' trajectories that remain inside the cavity during the
time period $2 \tau_{\rm E}$ is exponentially small, 
$\propto \exp(-\tau_{\rm E}/\tau_{\rm
  D})$, hence the exponentially small weak localization correction
if $\tau_{\rm E} \gg \tau_{\rm D}$.

Sofar we have considered the weak localization correction to the
transmission. What about other quantum interference effects, such
as the transmission fluctuations? Takane and Nakamura have extended
the semiclassical theory of Aleiner and Larkin to the conductance
variance $\mbox{var}\, G$, but for the limit $\tau_{\rm E} \to 0$
only.\cite{kn:takane1998} They found $\mbox{var}\, G = 2
(P_{\rm L} P_{\rm R})^2$, in agreement with predictions from
random matrix theory.\cite{kn:beenakker1997}
Just as the semiclassical theory for the
weak localization correction $\delta G_{\rm RL}$ was simpler than the
semiclassical theory for $\delta G_{\rm LL}$ (see above), the semiclassical 
theory of transmission fluctuations takes its simplest form if applied
to the covariance of reflection from the right contact $G_{\rm RR}$ and
reflection from the left contact $G_{\rm LL}$. Of course, unitarity 
implies $\mbox{var}\, G = \mbox{cov}\, (G_{\rm LL},G_{\rm RR})$. 
In Ref.\ \onlinecite{kn:takane1998}, only one contribution to the
conductance covariance $\mbox{cov}\, (G_{\rm LL},G_{\rm RR})$ was
considered. This contribution corresponds to the four trajectories
shown in Fig.\ \ref{fig:1}b. For these four trajectories, 
a minimal dwell time $2 \tau_{\rm E}$ is required: classical trajectories
originating from each of the openings need to diverge and 
reunite, each of which takes a time $\tau_{\rm E}$. Hence, following
the same phase space arguments as for the weak localization
correction, we anticipate that their contribution to the conductance
fluctuations depends on the Ehrenfest times as
\begin{eqnarray}
  [\mbox{cov}\, (G_{\rm LL},G_{\rm RR})]^{1/2}
  &=& [\mbox{var}\, G]^{1/2} \propto
  e^{-\tau_{\rm E}/\tau_{\rm D}} \nonumber \\ && \ \ \mbox{if $\tau_{\rm E} \gg
\tau_{\rm D}$}.
\end{eqnarray} 
However, there may be contributions to the conductance fluctuations
other than that of Fig.\ \ref{fig:1}b. (This possibility can not
be excluded on the basis of Ref.\ \onlinecite{kn:takane1998}.) For
example, there may be trajectories that involve small-angle
intersections of three trajectories at the same
point.\cite{kn:tian2004b,kn:mueller2004} While it is not expected
that the inclusion of such trajectories undo the suppression of
the conductance fluctuations in the
limit $\tau_{\rm E} \gg \tau_{\rm D}$, they may have a non-negligible 
contribution for $\tau_{\rm E} \sim \tau_{\rm D}$.


A final note about the results presented in this section:
In the calculation performed in this section 
we have replaced dynamical functions by their averages.
For example, the probability that a wave packet will exit
through a given lead is (almost) $1$ or $0$ for short enough times,
depending on the classical dynamics, not $P_{\rm L}$ or $P_{\rm R}$.
Thus, we have actually estimated the {\em ensemble} average of the
weak localization correction. One can expect that if a system
is close to the classical limit there will be more fluctuations in 
the values of diffusons and cooperons. This may lead to large
deviations of the weak localization correction from that of systems 
with small Ehrenfest times, and to large classical conductance
fluctuations.\cite{kn:tworzydlo2004} In order to describe these
fluctuations quantitatively, one needs a sample-specific theory for 
the dynamics of the diffuson, which covers the times of order
$\tau_{\rm E}$ exactly. While this is an important theoretical 
problem, it is beyond the scope of this paper, and should not 
affect the eventual exponential suppression of quantum interference 
phenomena for large Ehrenfest times.

\section{Discussion}
\label{discussion}

In Sec.\ \ref{numerics}, we reported results of numerical simulations for
weak localization, conductance fluctuations, and shot
noise of the open quantum kicked rotator. The numerical simulations
for weak localization and shot noise are consistent with
an exponential suppression 
$\propto \exp(-\tau_{\rm E}/\tau_{\rm D})$, in quantitative agreement
with the semiclassical theory of Sec.\ \ref{ALcalculation}. 
The numerical simulations for conductance
fluctuations show a small increase with increasing Ehrenfest time,
not inconsistent with previous simulation data reported in the 
literature.\cite{kn:jacquod2004,kn:tworzydlo2004,kn:tworzydlo2004c}
Our simulations also give information on the minimal time at which
quantum effects occur. For weak localization and shot noise, these
times are $2\tau_{\rm E}$ and $\tau_{\rm E}$, respectively, 
consistent with semiclassical theory. For conductance fluctuations,
the onset time is less than half the onset time of weak localization.
(Note that it is possible that some contribution for weak localization
in {\em reflection} may appear after a time of $\tau_{\rm E}$.)

In order to explain the numerical simulations of shot noise,
weak localization, and
universal conductance fluctuations, the authors of Refs.\ 
\onlinecite{kn:tworzydlo2003,kn:jacquod2004,kn:tworzydlo2004,kn:tworzydlo2004c}
proposed a phenomenological alternative to the standard semiclassical
theory, referred to as `effective
random matrix theory'. Guiding principle for the effective
random matrix
theory is that quantum diffraction takes place between trajectories with
dwell times larger than $\tau_{\rm E}$ only; trajectories with dwell
time shorter than $\tau_{\rm E}$ build scattering states
with transmission (exponentially close to) $0$ or $1$ and do not 
contribute to shot noise, weak localization, or universal conductance
fluctuations. The importance of `quantum' trajectories is described by 
an effective number of `quantum channels' $N_{\rm q} \sim N
\exp(-\tau_{\rm E}/\tau_{\rm D})$. 
Following Silvestrov {\em et al.},\cite{kn:silvestrov2003}
it was then proposed that the scattering of `quantum channels' is 
described by random matrix theory as long as $N_{\rm q} \gg 1$. The
condition $N_{\rm q} \gg 1$ is generically met, even if 
$\tau_{\rm E} \gg \tau_{\rm D}$.

The
effective random matrix theory not only correctly predicts the
suppression of the ensemble-averaged 
shot noise at large Ehrenfest times --- shot noise is 
proportional to the number of channels $N$, which is replaced
by $N_{\rm q} = N \exp(-\tau_{\rm E}/\tau_{\rm D})$ in the effective 
random matrix theory ---, it 
also describes sample-specific deviations from the ensemble
average that arise from the classical dynamics.\cite{kn:tworzydlo2003} 
The effective
random matrix also has had remarkable success explaining other
observables that are proportional to the channel number $N$, such as the
density of transmission eigenvalues,\cite{kn:jacquod2005} as
well as the density of states in a chaotic cavity coupled to a
superconductor.\cite{kn:silvestrov2003} 
On the other hand, quantum-interference effects, such as 
weak localization and conductance fluctuations are
independent of $N$, and, hence, are predicted to be independent of the 
Ehrenfest time. Thus, for weak localization, the effective random 
matrix theory differs from the semiclassical theory, which
predicts a suppression $\propto \exp(-\tau_{\rm E}/\tau_{\rm D})$.

The effective random matrix theory and the semiclassical description
of transport not only disagree regarding the magnitude of the weak 
localization correction to the conductance, they also disagree
with regard to the minimal time required for quantum interference 
effects to occur. In the semiclassical theory, quantum interference 
requires a minimal wavepacket to be split {\em and} reunited,
which takes a minimal time $2 \tau_{\rm E}$. This is in contrast to
the effective random matrix theory, where quantum interference
is fully established already after a time $\tau_{\rm E}$. 
Interestingly, there is no difference between semiclassics and 
effective random matrix theory for shot noise: Not
being a quantum interference effect, shot noise only requires 
wavepackets to be split, which happens after a time $\tau_{\rm E}$
in both theories. 

These two differences between the semiclassical theory for weak
localization and the
effective random matrix theory are not unrelated. {\em Both} 
theories
are consistent with a fully classical description of electron
dynamics for the first time interval of length $\tau_{\rm E}$ 
after the electron has entered the cavity. Differences between
the two theories appear for electrons that escape from the
cavity in the second time interval of length $\tau_{\rm E}$
({\em i.e.,} for times between $\tau_{\rm E}$ and $2 \tau_{\rm E}$).
In the semiclassical theory, electrons that escape during this
interval behave quantum mechanically but do not contribute 
to weak localization. As explained in Sec.\ \ref{ALcalculation},
the escape of electrons between $\tau_{\rm E}$
and $2 \tau_{\rm E}$ is responsible for the suppression of weak 
localization with exponent $\exp(-\tau_{\rm E}/\tau_{\rm D})$
in the semiclassical theory.
In the effective random matrix 
theory, weak localization sets in as soon as the classical
description fails, one Ehrenfest time after the electron enters
the cavity; 
there is no time interval in which electron dynamics 
is neither fully classical nor described by random matrix theory. 

A `microscopic theory' supporting the effective random matrix theory
and its prediction of Ehrenfest-time independent weak localization and
conductance fluctuations appeared recently.\cite{kn:whitney2005} This
theory is based on the assumption that electron dynamics in the
quantum trajectories ({\em i.e.,} trajectories with dwell times
larger than $\tau_{\rm E}$) is fully ergodic, with fully established
quantum interference corrections. Such an assumption violates the
semiclassical picture of weak localization, in which the quantum
correction only appears after a time $2 \tau_{\rm E}$. Hence, a
theory which tries to justify the effective random matrix theory must
provide an entirely new semiclassical model for weak localization, in 
which the quantum correction appears after one Ehrenfest time only. 

Although, at present, there is no theory that explains all simulation
data, we can conclude that our time-resolved simulation data for the
quantum interference corrections to the conductance are not
consistent with the effective random matrix theory. First, because the
effective random matrix theory predicts onset time $\tau_{\rm E}$
for weak localization, as well as an Ehrenfest-time independent
weak localization correction, both of which are ruled out by our
simulations. Second, because simulation results for weak localization
and conductance fluctuations are qualitatively different, whereas
the `effective random matrix theory' predicts equal onset times and
Ehrenfest-time dependences for weak localization and conductance 
fluctuations. 

Of course, the question why numerical simulations for weak
localization and conductance fluctuations in the open quantum kicked
rotator are qualitatively different remains.
Clearly, this question can not receive a final
answer as long as there is no semiclassical theory for the 
Ehrenfest-time dependence of conductance fluctuations. Given
the difficulty of obtaining accurate numerical results in 
the regime $\tau_{\rm E} \gg \tau_{\rm D}$, such a semiclassical
theory needs to include the parameter range 
$\tau_{\rm E} \sim \tau_{\rm D}$ if a valid comparison with the
numerical simulations is to be made.


\acknowledgments

We would like to thank C.\ Beenakker, S.\ Fishman, H.\ Schomerus,
P.\ Silvestrov, and D.\ Ullmo for discussions. 
This work was supported by the NSF
under grant no.\ DMR 0334499 and by the Packard Foundation.

\begin{appendix}

\section{Transmission and reflection in terms of diffusons}

In this Appendix the transmission and reflection through a quantum dot
are derived in terms of lead integrals of a diffuson.
Similar expressions were derived previously by Takane and 
Nakamura.\cite{kn:takane1997} However, their expressions are
ambiguous with respect to the 
exact locations of the cross sections in the leads.
In our approach, all ambiguities
are resolved.

To calculate the total transmission and reflection in terms of diffusons, it
is useful to obtain expressions for scattering matrix elements 
using the retarded Green function of the cavity.
Exact expressions of this type were derived in Refs.\
\onlinecite{kn:stone1988,kn:baranger1989}. The leads have
a uniform cross section, so that the lead wavefunction can be decomposed
into free waves along the lead (denoted by $x$ coordinates)
and a basis of transverse wavefunctions $\chi_a (y)$. 
The same decomposition can also be used for the 
retarded Green function (at the Fermi energy), and one defines
\begin{equation}
{\cal G}^{+} ({\bf r}_1, {\bf r}_2 ) = \sum_{mn} {\cal G}^{+}_{mn}
(x_1, x_2) \chi_m (y_1) \chi^{*}_n (y_2), \label{eq:G+}
\end{equation}
for coordinates ${\bf r}_1$ and ${\bf r}_2$ in the leads.
We use the convention that different leads are assigned different
modes, so that Eq.\ (\ref{eq:G+}) remains meaningful if ${\bf r}_1$ 
and ${\bf r}_2$ are in different leads. 

Using the asymptotics of ${\cal G}^+$ and of the scattering states, the Green
function in the leads can be written in terms of scattering matrix
elements $S_{mn}$. One finds\cite{kn:stone1988}
\begin{multline}
\label{leadG1}
{\cal G}^{+}_{mn} (x_1, x_2) =  - \frac{i}{v_n} \left[ \delta_{mn} e^{i k_n (x_1-x_2)} \rule{0pt}{17pt}\right. \\ + \left.  S_{mn} \sqrt{\frac{k_n}{k_m}}  e^{- i k_n x_2 - i k_m x_1} \rule{0pt}{17pt}\right], 
\end{multline}
for $x_1>x_2$ with $x_{1,2}$ in the left lead, and to
\begin{equation}
\label{leadG2}
{\cal G}^{+}_{mn} (x_1, x_2) =  -\frac{i}{v_n} S_{mn} \sqrt{\frac{k_n}{k_m}} e^{i k_m x_1 -i k_n x_2},
\end{equation}
for $x_1$ in the right lead while $x_2$ is in the left lead.
(In the following we use units where $\hbar=1$.)
For the following discussion it is important to emphasize the condition
$x_1>x_2$, which appears when both $x_1$ and $x_2$ are in the same lead.
One of the steps in the derivation of Eq. (\ref{leadG1})
involves an integral over a finite domain (bounded by the cross-section $x=x_2$),
where the integrand is proportional
to $\delta ({\bf r}_1 - {\bf r}_2)$. This integral will give different results
depending whether ${\bf r}_1$ is in the domain of integration ($x_1>x_2$)
or not ($x_1<x_2$). 
This leads to the inequality in Eq. (\ref{leadG1}).
Later, when we write ${\cal G}^{+}_{mn}$ in terms of
cross section integrals, two different cross sections, at $x_1$ and $x_2$,
are obtained.

To compute the conductance coefficients $G_{\rm LL}$ and $G_{\rm LR}$,
the absolute value squared of the matrix elements is needed. It can be 
expressed in terms
of Green function with the help of
\begin{multline}
\label{derivG}
 \left( \frac{\partial}{\partial x_1} - \frac{\partial}{\partial x_4} \right) \left( \frac{\partial}{\partial x_2} - \frac{\partial}{\partial x_3} \right)  \\ \times {\cal G}^{+}_{mn} (x_1,x_2) \left. \left[ {\cal G}^{+}_{mn} (x_4,x_3)\right]^{*} \right|_{\begin{subarray}{l} x_1=x_4 \\ x_2=x_3 \end{subarray}}   =  \\ \left\{ \begin{array}{ll} 4m^2 |S_{mn}|^2  & \mbox{ if } x_1 \in R, x_2 \in L \\
 4m^2 \left( \delta_{mn} - |S_{mn}|^2 \right) &  \mbox{ if } x_1,x_2 \in L, x_1>x_2. \end{array} \right.
\end{multline}
The conductance coefficients $G_{\rm LL}$ and $G_{\rm LR}$ are then
calculated using Eq.\ (\ref{conductances}). One can use
Eq. (\ref{derivG}) to simplify the summations,
since the lead modes enter into ${\cal G}_{mn}^{+}$ via cross section integrals
over the leads. One then uses
\begin{equation}
\sum_{n=1}^N \chi_n^* (y_1) \chi_n (y_4) \simeq \delta_{\lambda_F} (y_1-y_4).
\end{equation}
Replacing this approximate finite width delta function by an 
exact one leads, after a
straightforward calculation, to
\begin{align}
\label{leadintforrt}
  G_{\rm RL} & =  \frac{1}{4 m^2} \int_{C_{\rm R}} dy_1 \int_{C_{\rm L}'} d y_2  \left( \frac{\partial}{\partial x_1} - \frac{\partial}{\partial x_4} \right) \nonumber \\ & \times \left( \frac{\partial}{\partial x_2} - \frac{\partial}{\partial x_3} \right) \left. {\cal G}^{+} ({\bf r}_1,{\bf r}_2)\left[ {\cal G}^{+} ({\bf r}_4,{\bf r}_3)\right]^{*} \right|_{\begin{subarray}{l} {\bf r}_1={\bf r}_4 \\ {\bf r}_2={\bf r}_3 \end{subarray}} \\
  G_{\rm LL} & =   -\frac{1}{4 m^2} \int_{C_{\rm L}} dy_1 \int_{C_{\rm L}'} d y_2  \left( \frac{\partial}{\partial x_1} - \frac{\partial}{\partial x_4} \right) \nonumber \\ & \times \left( \frac{\partial}{\partial x_2} - \frac{\partial}{\partial x_3} \right) \left. {\cal G}^{+} ({\bf r}_1,{\bf r}_2)\left[ {\cal G}^{+} ({\bf r}_4,{\bf r}_3)\right]^{*} \right|_{\begin{subarray}{l} {\bf r}_1={\bf r}_4 \\ {\bf r}_2={\bf r}_3 \end{subarray}} \label{leadint2}.
\end{align}
This expressions differ from those in Ref.\ \onlinecite{kn:takane1997}
in that the integrals for the reflection are along separate cross sections,
with the cross section $C_{\rm L}$ closer to the cavity than $C_{\rm L}'$.

On each of the cross sections there are two points which are almost
identified: ${\bf r}_1 \simeq {\bf r}_4$ and ${\bf r}_2 \simeq {\bf r}_3$.
Therefore, it is natural the use the following Fourier transform~\cite{kn:aleiner1996,kn:takane1997}
\begin{multline}
\label{kd}
{\cal G}^+ ({\bf r}_1, {\bf r}_2) {\cal G}^- ({\bf r}_3, {\bf r}_4) = \int \frac{d {\bf p}_1}{(2 \pi)^2} \int \frac{d {\bf p}_2}{(2 \pi)^2} e^{i {\bf p}_1 \cdot ({\bf r}_1 - {\bf r_4})} \\  \times e^{i {\bf p}_2 \cdot ({\bf r}_3 - {\bf r_2})}  K^{\cal D} ({\bf p}_1, {\bf R}_1; {\bf p}_2, {\bf R}_2 ),
\end{multline} 
where ${\bf R}_1 = ({\bf r}_1+ {\bf r}_4)/2$ and ${\bf R}_2 = ({\bf r}_2+ {\bf r}_3)/2$.
Note that ${\bf R}_1$ and ${\bf R}_2$ are on the cross sections of the lead.
Energy conservation can be used to simplify this probability density further.
This is done by defining
\begin{multline}
\label{defdiff}
K^{\cal D} ({\bf p}_1, {\bf R}_1; {\bf p}_2, {\bf R}_2 )  =   \frac{2 \pi}{\nu} \delta \left( E_F - \frac{p_1^2}{2 m} \right) 
\delta \left( E_F - \frac{p_2^2}{2 m} \right)  \\  \times  {\cal D}  ({\bf n}_1, {\bf R}_1; {\bf n}_2, {\bf R}_2 ).
\end{multline} 
Substitution of (\ref{kd})
and (\ref{defdiff}) in (\ref{leadintforrt}) will lead to the 
expressions for $G_{\rm RL}$ and $G_{\rm LL}$ in terms of the
diffuson. The calculation is
straightforward. Some care is needed only when one determines the boundary
conditions. For $G_{\rm RL}$, only diffusions that go into the cavity
can get from the left lead into the right lead. 
These diffusons can arrive to the right lead only if they originate 
from the cavity. These considerations result in the Theta functions in Eq. (\ref{totT}).
The calculation for $G_{\rm LL}$ is very similar. The only difference is that 
both incoming and outgoing diffusons will cross the inner ($C_{\rm L}$) cross section.
However, {\em all} the incoming 
probability flux from $C_{\rm L}'$ must cross $C_{\rm L}$. This incoming
flux just cancels the $N_{\rm L}$ factor in (\ref{leadint2}). The
remaining contribution results in Eq. (\ref{totR}). 
\end{appendix}



\end{document}